\documentclass[]{imag-ms-template}
\usepackage[utf8]{inputenc}
\usepackage[T1]{fontenc}
\usepackage{svg}
\usepackage{subcaption}
\usepackage{booktabs}
\usepackage{float}
\usepackage{multirow}
\usepackage{graphicx}
\usepackage{adjustbox}
\usepackage{amsmath}

\DeclareMathOperator*{\argmin}{arg\,min}

\title{Automatic sub-bundle tractometry reveals localized microstructural variations in complex white matter topographies}

\author{Nathan Decaux$^{1}$, Jean-Charles Roy$^{2,5}$, Thomas Desmidt$^{3,4}$,\\Marie-Laure Paillere Martinot$^{6,7}$, Gabriel Robert$^{1,2}$, Julie Coloigner$^{1\ast}$\\
{\footnotesize $^{1}$Univ Rennes, INRIA, CNRS, INSERM, IRISA UMR 6074, Empenn ERL U 1228, 35000 Rennes, France.}\\
{\footnotesize $^{2}$Adult University Psychiatry Department, Guillaume Régnier Hospital, Rennes, France.}\\
{\footnotesize $^{3}$Université de Tours, INSERM, Imaging Brain \& Neuropsychiatry iBraiN U1253, 37032, Tours, France.}\\
{\footnotesize $^{4}$CHU de Tours, INSERM, CIC1415, 37000 Tours, France}\\
{\footnotesize $^{5}$Centre for Population Neuroscience and Stratified Medicine, Charité Universitätsmedizin Berlin, Berlin, Germany}\\
{\footnotesize $^{6}$INSERM U1299 "Developmental Trajectories \& Psychiatry", Centre Borelli, UMR9010 ;}\\
{\footnotesize  Dept mathematics, ENS Paris-Saclay, Univ Paris Saclay, Paris, France.}\\
{\footnotesize $^{7}$AP-HP Sorbonne Université, Department of Child and Adolescent Psychiatry, Pitié-Salpêtrière Hospital, Paris.}\\
{\footnotesize $^\ast$Correspondence: julie.coloigner@irisa.fr}
}

\begin{document}

\maketitle

\begin{abstract}

Conventional tractometry averages white matter microstructure into a single profile, masking intra-bundle spatial heterogeneity and diluting localized alterations. We propose an automatic sub-bundle tractometry framework using Fréchet-based hierarchical clustering to decompose bundles into geometrically coherent sub-bundles, resolving complex fanning configurations without empirical thresholds. Multi-compartment metrics (fractional anisotropy, FA; isotropic free-water fraction, IFW) are projected onto these clusters. 

We evaluated this framework in $140$ participants across early-life (ELD) and late-life (LLD) depression cohorts, alongside age-matched healthy controls. We assessed performance based on microstructural homogeneity, sensitivity to age, and clinical group differences. 

Proposed sub-bundle metrics were significantly more homogeneous than classical full bundle profiles ($p < 0.001$). Our approach showed enhanced sensitivity, detecting more age-correlated bundles in elderly controls ($11$ vs. $5$) and more clinical group differences in LLD (FA: $6$ vs. $4$; IFW: $23$ vs. $17$) with larger effect sizes. In ELD, it improved the detection of subtle FA differences between remitted and resistant individuals ($2.9\%$ vs. $1.8\%$). Gains were particularly pronounced in LLD, reflecting the framework’s ability to resolve fanning tracts susceptible to brain aging. By preserving spatial heterogeneity, this automatic decomposition provides a robust foundation for large-scale clinical tractometry.
\end{abstract}

\maketitle

\section{Introduction}

The characterization of white matter microstructure is crucial for advancing our understanding of brain connectivity and the mechanisms underlying neurological disorders~\citep{koshiyama2019white}. Diffusion MRI (dMRI) provides non-invasive access to white matter tissue properties. Fractional anisotropy (FA) metric derived from diffusion tensor model (DTI), is widely used as a proxy for microstructural integrity~\citep{lebihan2001diffusion,wijtenburg2012relationship}.

% More recently, more complex models, called multi-compartment models (MCMs), have been developed to provide more specific markers of axonal degeneration and neuroinflammation~\citep{sumra2025regional,panagiotaki2012compartment,kraguljac2022neurite,aronica2022association}.

% %% Tractography
% Tractography, on the other hand, allows the reconstruction of white matter fibers and the identification of anatomically defined set of fibers called bundles. By projecting microstructural metrics onto these bundle streamlines, the voxel-wise space of microstructural metrics is thus transformed into a tract-based space. 

More recently, multi-compartment models (MCMs) have been developed to provide biologically specific microstructural markers by decomposing the diffusion signal into multiple tissue compartments~\citep{panagiotaki2012compartment}. These models provide specific markers of neuroinflammation such as the isotropic free-water fraction (IFW)~\citep{hedouin2024microstructural,chang2025free}, and can disentangle multiple fiber populations with different directions within a single voxel, enabling fiber-specific FA estimates in crossing-fiber regions~\citep{witt2025tractspecific,mishra2015toward}. Exploiting the full specificity of these metrics, however, requires accurate spatial localization of the underlying fiber pathways, which tractography supplies. Through the reconstruction of white matter pathways, fibers can be grouped into anatomically defined bundles~\citep{wasserthal2018tractseg}. 
% Tractography reconstructs white matter fiber pathways and groups them into anatomically defined bundles. By projecting microstructural metrics onto these bundle streamlines, the voxel-wise space of diffusion measurements is transformed into a tract-based representation amenable to along-fiber analysis.

%% Tractometry
Tractometry consists in the analysis of the microstructural properties along these white matter bundles. In the conventional tractometry framework, each white matter bundle is represented by a single streamline, commonly a centroid streamline. This trajectory provides a simplified description of the bundle’s microstructure by aggregating microstructural measurements from the nearest points along the tract~\citep{tones2005pasta,corouge2004towards,cousineau2017testretest,chandio2023buan}. The resulting representation, called a $\textit{bundle profile}$, consists of a discrete set of points that capture microstructural values at specific locations along the pathway. These individual bundle profiles can then be compared between subjects, allowing for the identification and characterization of population-level differences in white matter microstructure through statistical analyses performed at the corresponding profile points \citep{yeatman2012tract,joo2025alongtract,zheng2026agfstractometry}.

% blabla limites single line
A key advantage of reducing a bundle to a single bundle profile is its ability to reduce the influence of measurements associated with aberrant streamlines, which may result from tractography reconstruction errors or inaccuracies in bundle segmentation~\citep{chamberland2019dimensionality}. However, reducing a bundle to a single bundle profile implicitly assumes uniform microstructural properties across its cross-section. This assumption is violated in many anatomically complex bundles with crossing fibers or fanning~\citep{kruper2021evaluating}, and as a result, group differences localised to a subset of streamlines may be diluted or entirely missed. In this regard, ~\citet{chandio2023buan} suggested to group bundle streamlines into multiple clusters and perform separate tractometry profiles onto these sub-bundles. However, the proposed clustering, based on the distance between streamlines, requires a fine-tuning of distance thresholds for each individual bundle. Moreover, the benefit compared to an analysis of the full bundle profile has not been assessed. An alternative yet complementary approach consists of clustering streamlines across subjects based on a whole-brain atlas~\citep{zheng2026agfstractometry,zhang2018anatomically} clustered with spectral embedding distances. By enabling sub-bundle profiling, this framework has been shown to improve the detection of localized microstructural abnormalities. Nevertheless, its effectiveness is highly contingent on the quality of the atlas, since the clustering process is conducted at the whole-brain level rather than within individual bundles.

% Another challenge in tractometry is to consistently associate streamline points with the bundle profile points across subjects. Early approaches partitioned each streamline into $k$ equal-length segments and mapped the points of each segment to the corresponding position along the profile~\citep{yeatman2012tract}. This strategy is, however, sensitive to streamline length variability, which is common in anatomically complex bundles. Subsequent methods instead assigned each streamline point to its nearest neighbour on the profile~\citep{chandio2020bundle}, but this approach struggles with associations in arc-shaped bundles.

To address the aforementioned limitations, we introduce a framework based on automatic clustering directly within a pre-existing anatomical bundle atlas, using strategies specifically designed to discriminate geometric variability such as crossing or fanning patterns. 
% Furthermore, we propose a novel point-level association strategy that combines spatial proximity with normalized arc-length information along the bundle. 
% We hypothesis that the combination of these two contributions  would provide an accurate association between streamline points and representative trajectories while preserving computational efficiency. 

% blabla microstructure et depression
A robust tractometry framework is particularly consequential in psychiatric research, where white matter alterations are typically subtle, spatially heterogeneous, and distributed across multiple pathways. Microstructural abnormalities have been frequently reported in fronto-limbic and association tracts~\citep{mesbah2023association,liao2012dysfunction} that shows large fanning at their cortical terminations. Large-scale meta-analyses have confirmed subtle but widespread white matter alterations in major depressive disorders~\citep{schmaal2020enigma}, yet the observed effects appear to be largely driven by medication status and comorbidities~\citep{jiang2017microstructural,xu2023metaanalysis}. While these analyses focused on FA, the use of MCM has revealed more specific alterations associated with depression, notably an increase in the isotropic free-water fraction (IFW)~\citep{langhein2022association,li2025white,bergamino2024distinguishing,vandeloo2022freewater,cao2025association,cox2016ageing}. However, effect sizes are still modest and the within-bundle spatial distribution of alterations remains poorly characterised~\citep{hedouin2024microstructural}. Therefore depression is an adequate framework to test for the sensibility of our approach. 

This is particularly relevant for late-life depression, where pathological changes intersect with brain aging, which is known to affect white matter in a non-uniform manner that preferentially impacts frontal and superficial regions~\citep{raz2006differential} where bundles are highly fanned.

%J'ai enlevé bergamino2015applying, car c'est du DTI corrigé FW 
We evaluated the proposed method against classical approaches by examining the microstructural heterogeneity of microstructure metrics (FA and IFW) within anatomically defined white-matter bundles. We further assessed its ability to detect group differences for these metrics and its sensitivity to subject age. This framework was validated on two independent cohorts of patients with depressive disorders. The first cohort comprised younger individuals with early-life depression (ELD), including both remitted and treatment-resistant phenotypes, while the second consisted of elderly individuals with late-life depression (LLD). In both cohorts, analyses were performed relative to age-matched healthy control groups.

\section{Methods}

\begin{figure}
    \centering
    \includegraphics[width=0.99\textwidth]{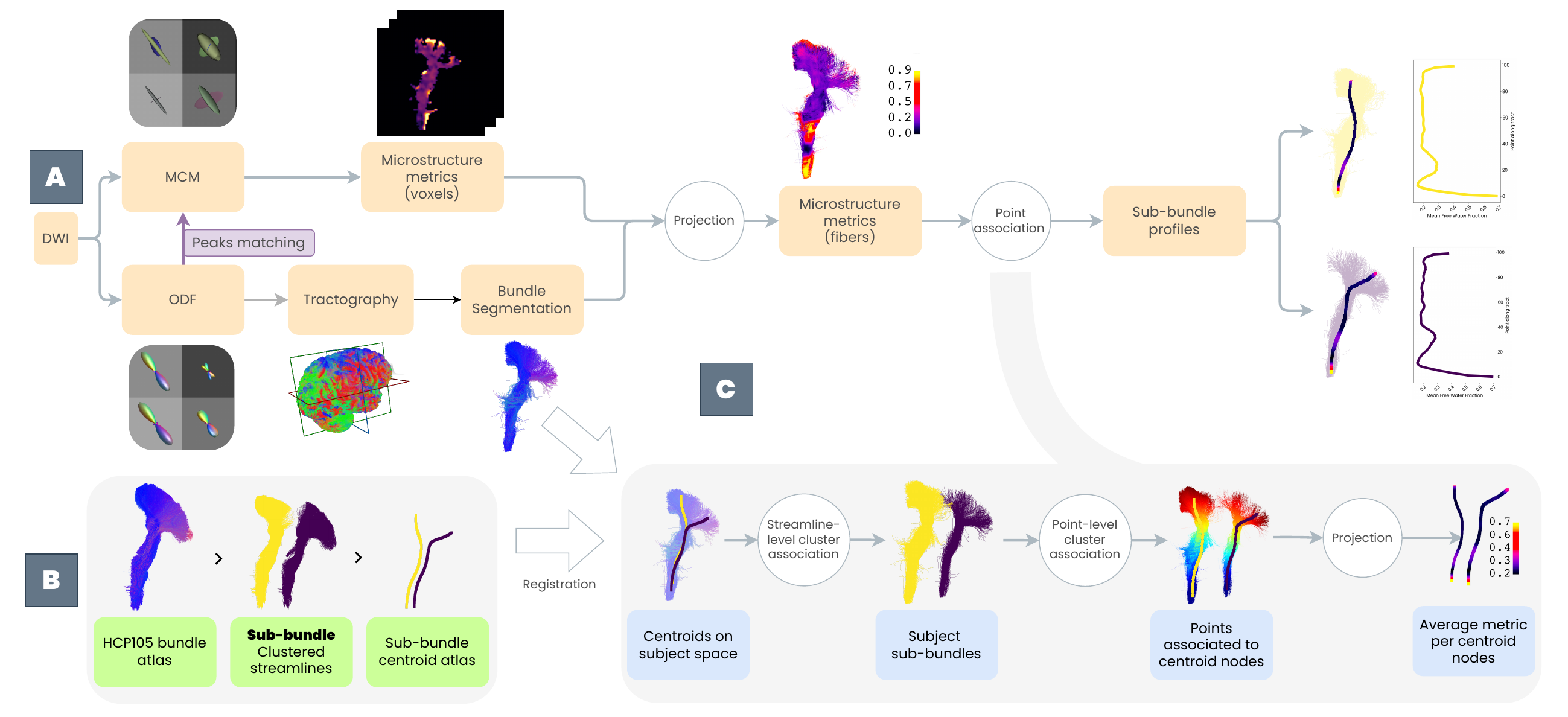}
    \caption{Overview of the proposed tractometry framework, illustrating key steps from diffusion MRI preprocessing to bundle profile analysis. \textbf{A. } Subject-level processing, from raw DWI signal to the projection of microstructure metrics onto identified tracts. \textbf{B. } Tractometry atlas construction and sub-bundle generation. \textbf{C. } Association steps between subject streamlines and atlas clusters, enabling the aggregation of microstructural metrics at anatomically corresponding locations along the bundles for subsequent statistical analysis.
    }
    \label{fig:pipeline}
\end{figure}

The overview of the proposed tractometry framework is illustrated in Figure \ref{fig:pipeline}. 

\subsection{Data and clinical context}
\subsubsection{Participants and clinical assessment}

Two independent cohorts of patients suffering from depression were included in this study. All participants provided written informed consent. Demographic and clinical characteristics of both cohorts are summarised in the supplementary materials (S1). Depression severity was assessed using the Montgomery-\r{A}sberg Depression Rating Scale (MADRS)~\citep{montgomery1979new}. All methods were performed in accordance with the guidelines of the Declaration of Helsinki.

\paragraph{ELD cohort}
It is a cross-sectional study that recruited $74$ participants ($25$ patients with remitted depression, $25$ patients with treatment-resistant depression (TRD)~\citep{holtzmann2016quelle}, and $24$ healthy controls (HC); $52$ females, $22$ males; mean age $26.1 \pm 5.6$ years) between July 2020 and January 2023 at Rennes University Hospital, France. Patients, aged 18-35 years, were recruited by their treating psychiatrists. Healthy controls were recruited via local advertisements and matched for age, sex, and education. TRD was defined as a lack of response to at least two antidepressant trials of adequate dosage and duration ($\geq$ 1 month). Remission was defined as a MADRS score $<$7 for $\geq$ 3 months. Exclusion criteria included substance use, bipolar, or developmental disorders; electroconvulsive therapy within six months; current somatic or neurological pathology; pregnancy; birth weight $<$800\,g; and standard MRI contraindications. All participants underwent a standardized psychiatric interview followed by an MRI scan on the same day. Treatment history was evaluated using the antidepressant treatment history Form, and alcohol consumption over the last month with the timeline followback method. The study was approved by the ethics committee CPP Ile-de-France (IDR-RCB 2013-A00662-43).

\paragraph{LLD cohort}
The LLD cohort recruited 60 participants ($36$ patients and $24$ healthy age-matched controls; $44$ females, $16$ males; mean age $74.9 \pm 6.0$ years) from the old-age psychiatry centres of Rennes ($39$ participants) and Tours ($21$ participants) between October 2019 and April 2022. Inclusion and exclusion criteria were assessed during a psychiatric interview conducted by a trained geriatric psychiatrist. Inclusion criteria were age $\geq60$ years with DSM-5 and Mini-International Neuropsychiatric Interview~\citep{sheehan1994mini} criteria for major depressive disorder. Non-inclusion criteria comprised DSM-5 criteria for major cognitive disorder and Mattis Dementia Rating Scale~\citep{mattis1976mental} $<125$; neurological diseases; inflammatory or mechanical diseases impeding motor activity; severe sarcopenia defined as a walk test $<$1\,m/s; extrapyramidal symptoms assessed by the Unified Parkinson's Disease Rating Scale-Part~III~\citep{disease2003unified}; having an antipsychotic prescription; high suicidality defined as a Clinical Global Impression Suicide Scale score~\citep{lindenmayer2003intersept}; legal protection or deprivation of liberty; and MRI contraindications. The study was approved by the relevant institutional review board (ID-RCB 2018-AO2643-52, NCT03807167).

Comparability in terms of age was assessed using Welch's ANOVA for ELD (p = 0.079) and Welch's $t$-test for LLD (p = 0.690) cohorts, with no significant differences between groups. In ELD, the MADRS score was significantly different between the resistant and the remission group ($26.2\pm5.9$ vs. $5.6\pm3.9, p \ll  0.001$, Welch's $t$-test).

\subsubsection{MRI scan acquisition}

At both sites (LLD - Rennes and Tours) and both samples, all participants underwent MRI in a 3T Siemens MR scanner (Magnetom Prisma, VE11C, Erlangen, Germany) with a 64-channel head coil. A whole brain T1-weighted MPRAGE image was acquired with repetition time (TR) = 1.9\,s, echo time (TE) = 2.26\,ms, inversion time (TI) = 900\,ms, flip angle = 9\textdegree, 1\,mm isotropic, field-of-view (FOV) = $256 \times 256$\,mm$^2$, 176 slabs. The multi-shell dMRI data were gathered with a CUbe and SPhere (CUSP) sequence~\citep{scherrer2012parametric} acquired on 72 slices using an interleaved slice acquisition, with the following parameters: slice thickness of 2\,mm, in-plane resolution = $2\,\text{mm} \times 2\,\text{mm}$, an acquisition matrix of $110 \times 110$, TR/TE = 5216/54.40\,ms, flip angle 90\textdegree, pixel bandwidth 1698\,Hz and an imaging frequency of 123.25\,MHz. The CUSP acquisition time was 6.42\,min. An additional $b_0$ volume with reversed phase encoding direction volume was also acquired with the same acquisition parameters for the distortion artifact correction. The specificity of this sequence lies in its 60 gradients that are placed on a sphere and a cube (i.e. with multiple gradient $b$-values ranging from 1000 to 3000\,s\,mm$^{-1}$). The goal of this particular gradient structure is to reduce the acquisition time compared to a regular multi-shell sequence while maintaining the quality of the resulting diffusion model. Moreover, the second interest of the CUSP sequence is in terms of image quality because the time echo of this diffusion sequence is less affected by high $b$-values, improving the signal-to-noise of the images.

\subsection{Subject diffusion processing}

The process from raw diffusion MRI data to tractometry analysis involves several key steps, including preprocessing, tractography, multi-compartment modeling and microstructural metric projection to streamlines. This pipeline is illustrated in the block A of Fig.~\ref{fig:pipeline}.

\subsubsection{Preprocessing}
Diffusion MRI preprocessing followed the protocol established by Hédouin et al. \citep{hedouin2024microstructural}, using the Anima toolbox, including motion and eddy-current correction with outlier replacement, distortion correction using the reverse phase-encoded b0 images and nonlocal-means filtering. A whole-brain mask was computed from the mean b0 image. Additional details are available in \citep{hedouin2024microstructural}.

\subsubsection{Tractography}
\paragraph{ODF}
Orientation distribution functions (ODFs) were estimated using the Multi-Shell Multi-Tissue Constrained Spherical Deconvolution (MSMT-CSD) algorithm~\citep{jeurissen2014multitissue} implemented in MRtrix3~\citep{tournier2019mrtrix}. Tissue response functions for white matter, gray matter, and CSF were estimated per subject and used to compute multi-tissue ODFs. Multi-tissue intensity normalization was applied to improve inter-subject comparability. To characterize local fiber populations, fiber bundle elements (fixels) were extracted per voxel following the approach by Raffelt et al.~\citep{raffelt2017investigating} and the MRtrix3 implementation.

\paragraph{Whole-brain tractography}
Probabilistic whole-brain tractography was performed using the iFOD2 \citep{tournier2010improved} algorithm with a target of 1 million streamlines, seeded from the whole-brain mask. Default MRtrix3 parameters were used, including curvature constraints and ODF amplitude cutoff. Minimum and maximum streamline lengths were set to exclude implausible tracts, and tracking was confined within the brain mask. All tractography outputs underwent basic quality check to verify streamline distribution and anatomical plausibility.

\subsubsection{Multi-compartment model}
In MCMs, the diffusion-weighted signal is expressed as the weighted sum of multiple compartments, each representing a distinct microstructural environment with specific diffusion properties~\citep{panagiotaki2012compartment}. Isotropic compartments account for free water or approximately spherical structures such as inter-axonal restricted water, while anisotropic compartments capture the directional diffusion observed within coherent fiber populations. Formally, the signal model decomposes the diffusion probability density into a mixture of isotropic and anisotropic components, with nonnegative compartment weights that sum to one. 

Assuming Gaussian compartmental diffusion, we use a MCM comprising two isotropic compartments, modeling free water and restricted water, together with a variable number of anisotropic compartments, each parameterized by a diffusion tensor. To robustly handle distinct crossing fibers, we used a variable number of anisotropic compartments, $k$, ranging from $0$ to $3$ based on the previously extracted number of fixels (prominent ODF peaks). This flexibility accommodates diverse regimes: no anisotropic compartment in CSF-dominated voxels, a single dominant compartment in coherent bundles (e.g., corpus callosum), and two or three compartments in complex crossing regions.

\paragraph{Derived microstructure metrics}
From the fitted model, we derive the following microstructural metrics: IFW from the isotropic compartment; iso-restricted (``trapped'') water fraction (IR); and, for each anisotropic compartment $j = 1,\ldots,k$, FA is computed from the corresponding diffusion tensor. 

\paragraph{Projection to tractography}
Voxelwise MCM metrics are interpolated onto streamline points. For the FA, a direction-dependent metric, each streamline point is associated with the anisotropic compartment that maximizes the absolute dot product between its principal orientation and the local streamline tangent.

\subsection{Tractometry atlas}

% \begin{figure}
%     \centering
%     \includegraphics[decodearray={0 1 0 1 0 1},width=0.99\textwidth]{figures/clustering_more_contrast.drawio.png}
%     \caption{Clustering of atlas streamlines into sub-bundle clusters. Left: Comparison of baseline vs $\text{Endpoint}$-based clustering methods for a crossing streamline bundle example. Right: Comparison of baseline vs distance-based clustering methods for a fanning bundle example. Top row: schematic illustration of the clustering methods. Bottom row: example bundles from the HCP atlas, with streamlines colored by cluster assignment (CC 3 and left CST bundles, respectively).
%     }
%     \label{fig:clustering}
% \end{figure}

Reference atlas is crucial to identify the bundles of interest to be extracted, as well as providing a reference geometry for the extraction of clusters and the alignment of individual fibers. The key steps involved in this process are illustrated in the block B of Fig.~\ref{fig:pipeline}.

\subsubsection{Streamline atlasing}

We build an HCP105 bundle atlas~\citep{jakob2018high}, comprising anatomically defined bundles reconstructed from $105$ subjects. For each HCP diffusion data, fiber bundles were semi-automatically obtained by \citet{wasserthal2018tractseg}, using fiber tracking and filtering, first by regions of interest (ROIs), then curated by an expert. The overall process used to segment these bundles is detailed in \citet{wasserthal2018tractseg} and the data is available online \citep{jakob2018high}. A total of 71 white matter bundles were investigated encompassing seven subdivisions of the corpus callosum (CC: rostrum, genu, rostral body, anterior midbody, posterior midbody, isthmus, and splenium), the anterior commissure (CA), association fascicles including the arcuate (AF), cingulum (CG), inferior occipito-frontal (IFO), inferior longitudinal (ILF), middle longitudinal (MLF), superior longitudinal I–III (SLF I–III), and uncinate (UF) fasciculi; projection tracts including the corticospinal (CST), anterior and superior thalamic radiations (ATR, STR), fronto (FPT) and parieto-occipital (POPT) pontine, optic radiation (OR), and fornix (FX), cerebellar peduncles including the inferior, middle, and superior (ICP, MCP, SCP), thalamic (T) pathways to prefrontal (PREF), premotor (PREM), precentral (PREC), postcentral (POSTC), parietal (PAR), and occipital (OCC) regions, and striato-cortical (ST) pathways to the same cortical targets plus fronto-orbital (FO) cortex. 

Bundles were then registered to a common space using an atlasing method, initially proposed by \citet{durantel2025riemannian}. It is a modified version of~\citet{guimond2000average} approach applied to diffusion data~\citep{suarez2012automated}. It iteratively builds a tensor atlas by registering each individual HCP tensor images onto a running reference and averaging the resulting unbiased atlases at each step. Key modifications include the use of diffeomorphic Stable Vector Fields (SVF) and the log-Euclidean framework~\citep{arsigny2006logeuclidean} for computing average transformations. The transformation field obtained by the atlasing method from each individual space to the atlas space are then applied to $71$ segmented fiber bundles of interest of each subject in order to all align them to the tensor atlas. For each bundle, all streamlines from all subjects were, then, merged into a single tractogram to preserve the variability of individual shapes. This merged tractogram served as the bundle atlas for subject fiber segmentation and subsequent clustering and cluster extraction. Transformations and related scripts use to generate the bundle templates are publicly available~\footnote{\url{https://zenodo.org/records/22641462}}.

\subsubsection{Bundle clustering}

\begin{figure}
    \centering
    \includegraphics[width=1\linewidth]{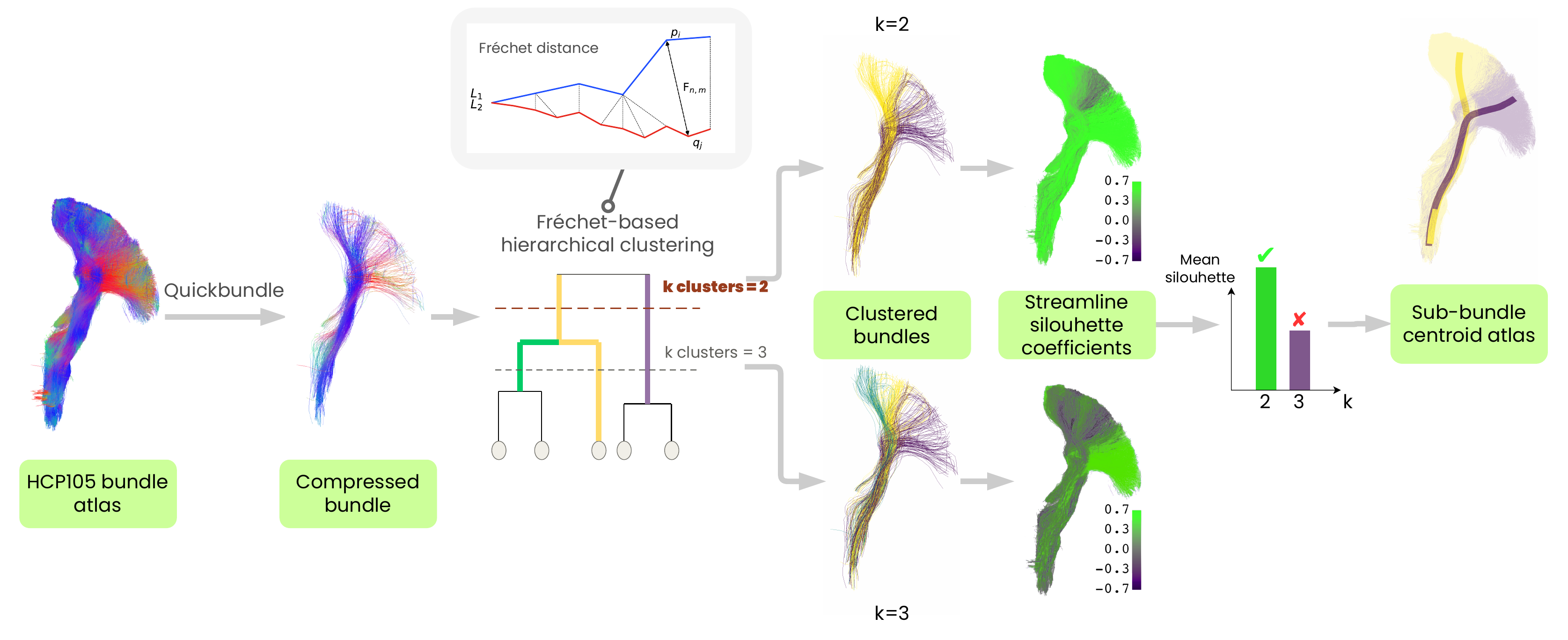}
     \caption{Fréchet-based sub-bundle clustering pipeline, illustrated on the left corticospinal tract (CST), related to block B in Fig.~\ref{fig:pipeline}. The bundle atlas (HCP105) is first compressed with QuickBundles ($5\,\mathrm{mm}$ threshold) to obtain centroids. Pairwise discrete Fréchet distances between centroids are then grouped by agglomerative hierarchical clustering with complete linkage, yielding candidate partitions for $k \in \{2,3\}$. For each candidate, per-streamline silhouette coefficients are computed, and the number of clusters maximizing the mean silhouette score is retained (here $k=2$). The selected clustering, with its cluster centroids, defines the reference sub-bundles used for subsequent tractometry.}
    \label{fig:frechet_clustering}
\end{figure}

%Clustering can be used to identify sub-groups of streamlines sharing similar geometry within a bundle. In conventional approach, a single centerline per bundle is extracted~\citep{chandio2020bundle}. However, this may be overly simplistic for bundles with complex internal organisation, such as those exhibiting lateralised sub-components or heterogeneous spatial topographies. To address this limitation, we propose a multiple centerline approach in which multiple clusters are extracted per bundle. The idea of further subdividing a bundle has already been explored, notably by Chandio et al. for the same purpose. 
Clustering can be used to identify sub-groups of streamlines with similar geometric characteristics within a bundle. In conventional approaches, a single representative centroid is typically extracted for each bundle~\citep{chandio2020bundle,cousineau2017testretest}. However, this representation may be insufficient for bundles exhibiting heterogeneous spatial topographies. We hypothesize that microstructural variations may be confined to specific spatial sub-region of a bundle, requiring these regions to be analyzed independently to avoid averaging out localized effects. Partitioning bundles has previously been explored by ~\citet{chandio2020bundle} to mitigate this issue. In this work, the proposed clustering strategy is based on a conventional pairwise distances between streamlines known as the average direct-flip distance (MDF), first introduced in the QuickBundles algorithm by ~\citet{garyfallidis2012quickbundles}. While computationally efficient, this approach relies on a distance threshold that must be tuned empirically for each bundle. Moreover, this average streamline distance is a suboptimal metric in certain configurations. Although it can discriminate globally distant streamlines or of different lengths, it fails to separate fibers that share a similar trajectory but diverge at one extremity. This causes streamlines with distinct fanning patterns to be incorrectly grouped together, obscuring the true spatial spread of the bundle. To address this issue, we investigate a strategy specifically designed to capture geometrically distinct sub-bundles, and is illustrated in Fig.~\ref{fig:frechet_clustering}.

Instead of MDF distance, we introduce pairwise streamline distances computed using the discrete $\text{Fréchet}$ distance \citep{eiter1994computing}, which measures the maximum pointwise distance along ordered point sequences. Formally, let $\mathrm{c}_a = (\mathrm{p_a}^{(1)}, \ldots, \mathrm{p_a}^{(L_a)})$ and $\mathrm{c}_b = (\mathrm{p_b}^{(1)}, \ldots, \mathrm{p_b}^{(L_b)})$ be two streamlines of $L_a$ and $L_b$ points respectively. Their discrete Fréchet distance $\mathrm{F}_{L_a, L_b}$ is computed by the recurrence:
$$\mathrm{F}_{i,j} = \max\!\bigl(d(\mathrm{p_a}^{(i)}, \mathrm{p_b}^{(j)}),\;\min(\mathrm{F}_{i-1,j},\,\mathrm{F}_{i,j-1},\,\mathrm{F}_{i-1,j-1})\bigr)$$
with $i \in \{1,\ldots,L_a\}$ and $j \in \{1,\ldots,L_b\}$, where $d(\cdot,\cdot)$ denotes the Euclidean distance. To reduce computational cost, an initial clustering step was performed using the QuickBundles algorithm~\citep{garyfallidis2012quickbundles} with a conservative distance threshold of $5,\mathrm{mm}$, to obtain a compressed bundle. The Fréchet distance was subsequently computed between the resulting cluster centroids rather than between all individual streamlines.

These centroids were then grouped using agglomerative hierarchical clustering with complete linkage applied to this Fréchet distance matrix~\citep{mullner2011modern}. A key advantage of this formulation is that it operates on a fixed number of clusters $k$ rather than a distance threshold, avoiding the need for per-bundle empirical tuning. The optimal number of sub-bundles, restricted to $k \in \{2,3\}$, was determined automatically by maximizing the average streamline silhouette coefficient~\citep{shahapure2020cluster}. For a given streamline $i$, the silhouette coefficient is defined as $s(i)=\frac{b(i)-a(i)}{\max(a(i),b(i))}$, where $a(i)$ denotes the average distance between $i$ and all other streamlines within the same cluster, and $b(i)$ corresponds to the minimum average distance between $i$ and streamlines belonging to the nearest neighboring cluster. The value of $k$ achieving the highest mean silhouette over all streamlines is retained as the optimal partition. Once the optimal clustering was identified, each original streamline was assigned to a final sub-bundle according to the label of its corresponding QuickBundles cluster from the initial grouping stage. Sub-bundle centroids are computed for each cluster using the standard centerline generation methodology~\citep{chandio2023buan,yeatman2012tract}.

\subsection{Subject-level bundle correspondence and tractometry}

Mapping subject streamlines onto the atlas-defined sub-bundles is a critical step for ensuring that microstructural profiles are comparable across individuals. This process involves three key stages: bundle segmentation, streamline-level cluster association, and point-level cluster association. First, the anatomical bundles of interest are identified within each subject's tractogram. Next, individual streamlines are associated with the corresponding atlas clusters. Finally, each point along the streamlines is mapped to a cluster-specific location, establishing a fine-grained correspondence with the atlas. Together, these steps enable the aggregation of microstructural measurements at anatomically consistent locations along the bundles, thereby facilitating meaningful inter-subject comparisons. This process is illustrated in the block C of Fig.~\ref{fig:pipeline}.

\subsubsection{Bundle segmentation}
Anatomical bundles were identified using BundleSeg~\citep{st-onge2023bundleseg} with the HCP105 bundle atlas as reference. FA map estimated from tensor atlas were non-linearly aligned to the FA map of each subject. Streamlines were, then, assigned to target bundles using distance-based matching provided by BundleSeg with a fixed radius of $8 mm$. This procedure produced a labeled set of streamlines for each bundle of interest and ensured consistent definitions across subjects.

\subsubsection{Streamline-level cluster association}

Sub-bundle centroid atlas was registered to the subject's native space using the same transformation employed in bundle segmentation.  Each subject streamline was assigned to its nearest atlas cluster based on point-wise Euclidean distance to obtain subject sub-bundles. Following the approach of Garyfallidis et al.~\citep{garyfallidis2012quickbundles}, this distance metric accounts for streamline orientation by evaluating both the original and reversed orientation, retaining the configuration that produces the highest similarity. The most similar orientation is then kept for subsequent point-level association.

\subsubsection{Point-level cluster association}

After streamline-to-cluster assignment, each point along a subject’s streamline must be matched to a corresponding point within the associated atlas cluster. Several methods have been proposed to perform this association, including nearest-neighbor matching based on Euclidean distance~\citep{chandio2020bundle} and correspondence based on index position along the streamline~\citep{yeatman2012tract}. However, both strategies exhibit important limitations (illustration in the supplementary material, Fig.~\ref{fig:association_method}). The minimum Euclidean distance can produce erroneous correspondences in highly curved or U-shaped bundles, while the projection based on index position does not necessarily reflect anatomical equivalence with different lengths or exhibiting local stretching. More sophisticated approaches have been proposed to address these limitations. For example, the RadTract framework introduced by Neher et al.~\citep{neher2023radiomic} trains a support vector machine (SVM) to learn a set of decision boundaries that partition streamline points according to their anatomical location, subsequently assigning each point to the nearest parcel in the learned feature space. While effective, this approach is computationally demanding, as a separate model must be trained for each sub-bundle.

%More advanced methods, such as the RadTract approach proposed by Neher et al.~\citep{neher2023radiomic}, train a support vector machine (SVM) to learn the hyperplane that best separates streamline points based on their position, and then assign points to the nearest parcel in this learned space. However, this method is computationally intensive as it must be retrained for each sub-bundle.

To overcome these limitations, we propose a position-weighted distance metric that integrates both spatial and longitudinal information along the fiber trajectory. The metric jointly penalizes geometric separation and mismatches in relative streamline position, thereby promoting correspondences between points and cluster points that are both spatially nearby and similarly located along the bundle. An illustration of the proposed association strategy is provided in the supplementary material (Fig.~\ref{fig:association_method}).

Given the subject streamline $s$ and the previously associated sub-bundle centroid $c$, let $\mathrm{p}_s^{(i)}$ denote the 3D coordinates of the $i$-th point ($i \in \{1,\ldots,L_s\}$) of $s$ with a length of $L_s$ points, and $\mathrm{p}_c^{(j)}$ the 3D coordinates of the $j$-th  ($j \in \{1,\ldots,L_c\}$) of the $c$ resampled to $L_c$ points. To jointly penalize spatial displacement and length position mismatch, each point is embedded into an augmented four-dimensional space by appending its normalized length coordinate, scaled by a weight set to the arc-length of the subject streamline $\alpha = \sum_{i=1}^{L_s-1} \left\|\mathrm{p}_s^{(i+1)} - \mathrm{p}_s^{(i)}\right\|_2$. The augmented space is then defined as : 
$$\tilde{\mathbf{p}}_s^{(i)} = \left(\mathrm{p}_s^{(i)},\; \alpha\,\frac{i-1}{L_s-1}\right), \qquad \tilde{\mathbf{p}}_c^{(j)} = \left(\mathrm{p}_c^{(j)},\; \alpha\,\frac{j-1}{L_c-1}\right).$$

The association distance is then the Euclidean distance in this augmented space
$$d_{\text{assoc}}(i, j) = \left\|\tilde{\mathbf{p}}_s^{(i)} - \tilde{\mathbf{p}}_c^{(j)}\right\|_2 = \sqrt{\left\|\mathrm{p}_s^{(i)} - \mathrm{p}_c^{(j)}\right\|_2^2 + \alpha^2\!\left(\frac{i-1}{L_s-1} - \frac{j-1}{L_c-1}\right)^{\!2}}.$$

This choice provides an automatic, scale-invariant calibration: the arc-length penalty term equal to $$\alpha\left|\frac{i-1}{L_s-1} - \frac{j-1}{L_c-1}\right|,$$whose maximum value $\alpha$ is exactly equal to the total length of the subject streamline, placing it on the same scale as the 3D spatial term. Consequently, a maximal normalized position mismatch $\left|\frac{i-1}{L_s-1} - \frac{j-1}{L_c-1}\right|=1$ contributes a penalty comparable to traversing the full fiber length, while neighboring points $\left|\frac{i-1}{L_s-1} - \frac{j-1}{L_c-1}\right| = 0$ are penalized almost exclusively by spatial distance. This balancing is automatic and requires no tuning, regardless of fiber size or resampling resolution. Correspondences are found efficiently for all streamline points simultaneously using a k-d tree built on the augmented cluster coordinates. This keeps computational cost similar to a standard nearest neighbor search while providing a more anatomically meaningful association that respects both spatial proximity and fiber trajectory.

Each streamline point is assigned to the cluster point with lowest association distance:
$$\phi(i) = \argmin_{j \in \{0,\ldots,L_c\}} d_{\text{assoc}}(i,\, j)$$

The set of streamline points attributed to cluster point $j$ is then:
$$\mathcal{S}(j) = \bigl\{\, \mathrm{p}_s^{(i)} \mid \phi(i) = j \,\bigr\}$$

\subsection{Tractometry analysis}

For each point $j$, the microstructural value was computed by averaging the metric values of all points in $\mathcal{S}(j)$, resulting in a single scalar value per point and per subject (see the last step of the block C of the Fig. 1). This procedure generated  sub-bundle profiles for each subject, bundle, and cluster. Our analyses were restricted to FA and IFW metrics, as these metrics provide direct biological insight into tissue alterations, with FA reflecting the loss of microstructural integrity associated with axonal degeneration and IFW capturing increases in extracellular free water linked to neuroinflammatory processes.
The cluster length $L_c$ was set to 50 points for all bundles. Empirically, this value provided an appropriate balance between spatial resolution, allowing the detection of localized group differences, and robustness to local anatomical variability and noise.

Because of inter-subject differences in bundle geometry and imperfections in spatial registration, the number of streamline points assigned to a given cluster point may vary across subjects, and some points may receive no assignment in certain cases. To ensure robust and consistent statistical comparisons, we identified, for each sub-bundle, the longest contiguous profile segment containing valid data for all subjects. Points lying outside this common segment were excluded from subsequent analyses.

\subsubsection{Bundle-level statistics}
To assess the quality of the sub-bundle approach, we evaluated several criteria for each bundle and microstructural metrics and compared them to the classical full bundle method.

\paragraph{Microstructural homogeneity}
%  First, a Welch's ANOVA was performed on the mean microstructural metric across sub-bundle clusters (one value per subject and cluster), testing whether the different clusters of a given bundle carry statistically distinct microstructural information. A significant result ($p < 0.05$) indicates that the clustering captures meaningful intra-bundle spatial heterogeneity. 

The first assumption is that meaningful clustering should capture microstructural heterogeneity within the bundle, as reflected by significant differences in the mean values of microstructural FA and IFW. To evaluate this hypothesis, within-cluster homogeneity was quantified by computing the standard deviation (SD) of microstructural values across all points and subjects assigned to each cluster, with lower SD values indicating greater homogeneity. For each subject and bundle, two summary measures were derived: the mean SD across all sub-bundles and the SD of the most homogeneous sub-bundle. The latter was compared with the mean sub-bundle SD using a paired $t$-test. In addition, both measures were contrasted with the SD computed over the classical full bundle configuration using Welch’s $t$-tests performed across bundles. All the above analyses were performed independently for each microstructural metric (FA, IFW) and method after residualization on confounders (age only).

\subsubsection{Along-tract statistics}
\label{sec:along_tract_stats}
Similarly to the \textit{Automated Fiber Quantification} (AFQ) approach~\citep{yeatman2012tract}, point-wise statistical tests were computed at each cluster point, producing a profile of test statistics along the tract for each bundle. To control for multiple comparisons across points, we applied a two-stage correction procedure. First, contiguous clusters of nominally significant points were identified using a cluster-based family-wise error rate (clusterFWE) permutation framework; second, only clusters containing at least one point whose corrected $p$-value fell below a point-level family-wise error threshold ($\alpha_{\text{FWE}}$) were retained as significant. This combined strategy ensures that reported effects are both spatially extended and include at least one point with strong individual-level evidence, thereby reducing the risk of reporting spatially diffuse but weak effects.

\paragraph{Age sensitivity}
Aging is a major factor affecting white matter microstructure, and thus a meaningful representation should be sensitive to age-related changes. To evaluate this, point-wise Pearson correlations between microstructural metrics and subject age were computed at each cluster point, restricted to healthy controls to avoid confounding by disease-related changes.

\paragraph{Group comparison}
To characterise the spatial distribution of group effects along the bundles, Welch's $t$-tests comparing clinical groups were computed at each point. We consider that a larger number of significant points within a bundle indicates a more sensitive representation, as it captures more extensive spatial patterns of group differences.

\section{Results}

Applying the sub-bundle clustering method to the 71 bundles of the HCP105 atlas, 58 bundles were divided into two sub-bundles and the remaining 13 into three. In the two sub-bundles setting, the average Fréchet distance from a streamline to others in its own sub-bundle is 36\% smaller than to streamlines in the nearest neighboring one (silhouette 0.36). For the 13 bundles split into three, this figure is 29\% (silhouette 0.29). This makes a total of $155$ sub-bundles to study. 

Since the assignment of a subject's streamlines to sub-bundles is based on the distance from the centroids, there is no guarantee that a subject's streamlines will be distributed across all sub-bundles. Among the $155$ sub-bundles, $41$ and $5$ were not assigned to every LLD and ELD subjects, respectively. The number of missing subjects ranges up to $16/60$ in LLD and $8/75$ in ELD. These sub-bundles are not included in the subsequent analysis.

However, in the ELD cohort, all bundles have at least one sub-bundle associated with all subjects. In the LLD cohort, only $8\//71$ bundles were missing subjects across their sub-bundles (left and right ILF, left and right FX, CA, left SLF I, right SLF II, and left SLF III) and were excluded from the \textit{Group comparison} (Sec.~\ref{ssec:group_comparison}) analysis. These specific bundles are illustrated in Fig.~\ref{fig:excluded_bundles}.

Among these bundles, only four (left ILF, right FX, CA, and left SLF III) lacked healthy controls in every sub-bundle, and were therefore also excluded from the \textit{Age sensitivity} (Sec.~\ref{ssec:age_sensitivity}) analysis.

\subsection{Microstructural homogeneity}

The proposed clustering method revealed substantial microstructural heterogeneity within white matter bundles, producing significantly different sub-bundle profiles across both cohorts and metrics. In the ELD cohort, significantly increased FA homogeneity was observed in 42 of 71 bundles, alongside increased IFW homogeneity in 47 out of 71 bundles. Similarly, in the LLD cohort, we found significantly increased homogeneity within 48 bundles for FA and 34 bundles for IFW out of 71.

\begin{figure}
    \centering
    \resizebox{\textwidth}{!}{
    \includegraphics[width=0.75\linewidth]{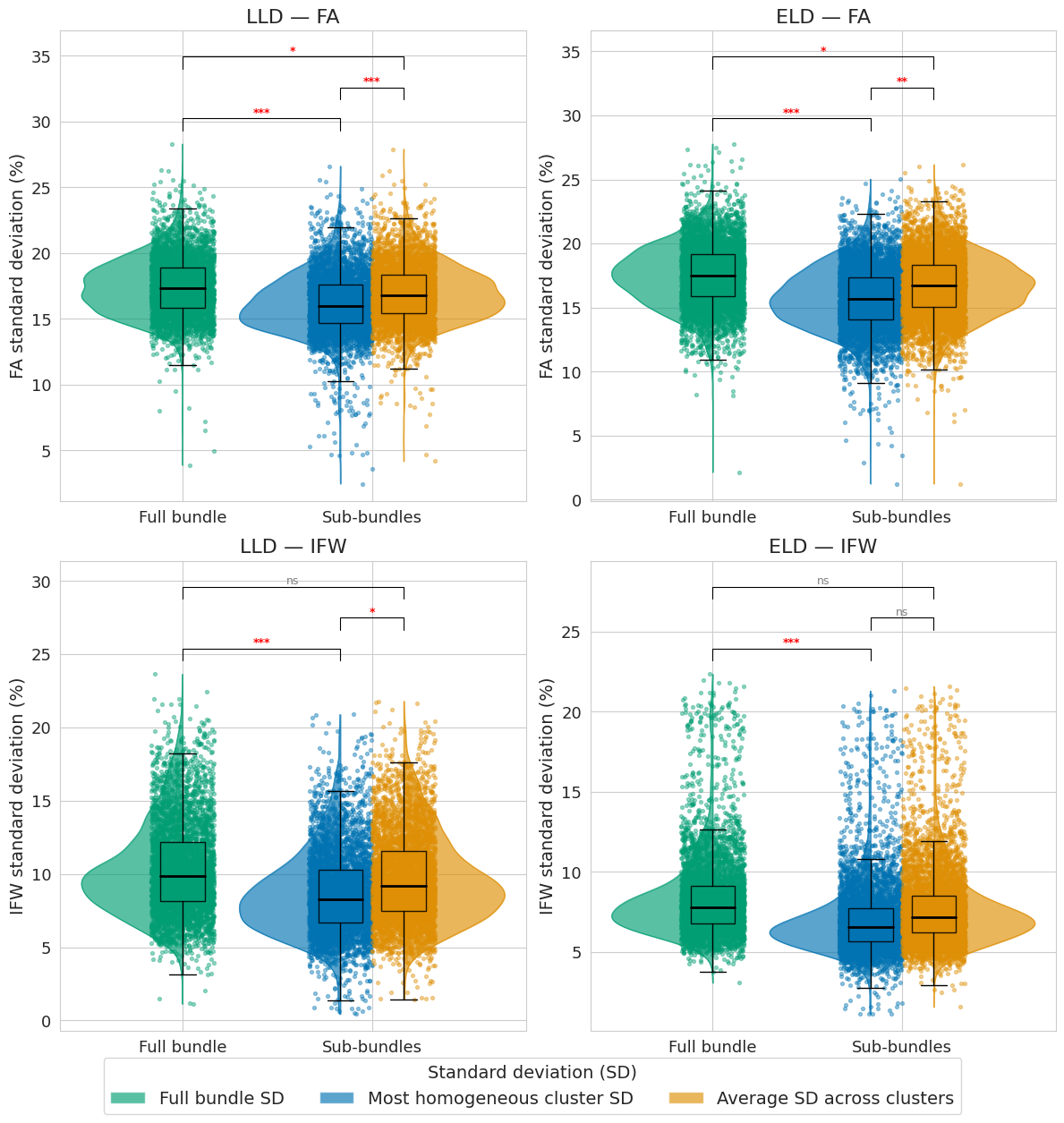}
    }
    \caption{Distribution of the mean within-cluster standard deviation (in metric percentage) for the full bundle and sub-bundle approaches, shown per cohort (LLD, ELD) and metric (FA, IFW). For the sub-bundle approach, two summaries are reported: the most homogeneous cluster (lowest SD) and the average SD across all clusters of each bundle. Statistical differences are tested between the full bundle approach and both sub-bundle summaries, as well as between the two sub-bundle summaries. Significance levels are indicated as * for $p < 0.05$, ** for $p < 0.01$, and *** for $p < 0.001$}
    \label{fig:std_best_centroid}
\end{figure}

Figure~\ref{fig:std_best_centroid} compares the within-cluster SD of microstructural metrics obtained with the proposed $\text{sub-bundle}$ clustering against the conventional full bundle analysis with only one centroid. The proposed approach produced significantly more homogeneous clusters than the full bundle method  ($p < 0.01$ to $p < 0.001$). When considering the average SD across all sub-bundles, the \text{sub-bundle} clustering reduced IFW variability from 8.4\% to 7.8\% in the ELD cohort and from 10.4\% to 8.7\% in the LLD cohort. Selecting the most homogeneous sub-bundle in each bundle resulted in an additional reduction in SD compared with the average across sub-bundles (p < 0.001 for all conditions). This reduction ranged from 0.7\% to 1.0\%.

\subsection{Age sensitivity}
\label{ssec:age_sensitivity}

\begin{figure}[h]
    \centering
    \includegraphics[width=0.99\linewidth]{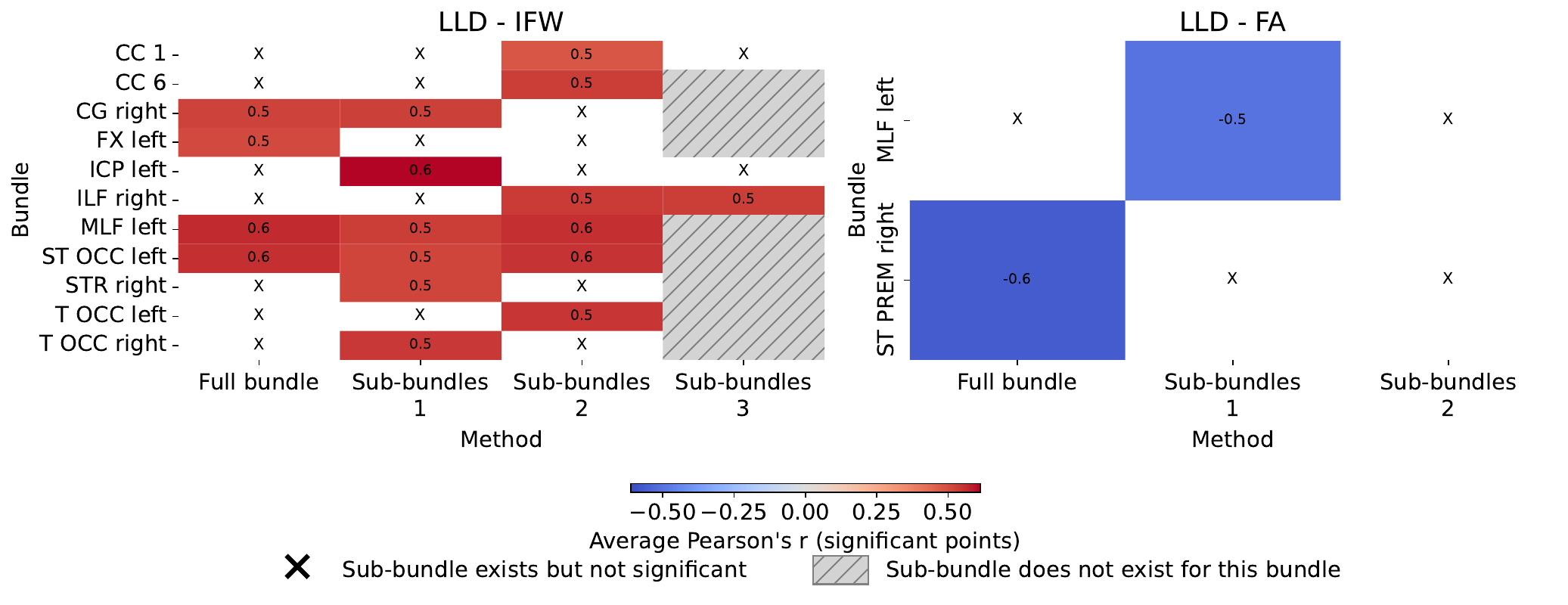}
    \caption{Correlation matrices between control subject age and microstructural metrics for the LLD dataset, per bundle or sub-bundle. Reported values are the average Pearson's $r$ over significant (multiple-comparison corrected) points only in the along-tract analyses. In sub-bundle settings, when significant points were detected for the same bundle in several clusters, values are reported separately with an arbitrary index (sub-bundle 1 or sub-bundle 2). Excluded bundles for analysis : left ILF, right FX, CA, and left SLF III.}
    \label{fig:age_correlation}
\end{figure}

Results of the along-tract (or tract-cluster) microstructural changes correlation with age in controls only is reported in the Fig.~\ref{fig:age_correlation}. In both approaches, similar correlations are detected in both dataset and metrics. In LLD, age is correlated positively correlated with IFW (pearson coefficients between $0.5$ and $0.6$) and negatively correlated with FA ($-0.5$ and $-0.6$). Overall, the sub-bundle clustering detected more significant ($11$ vs $5$) bundles than the standard full bundle analysis. The left FX in IFW and the right ST PREM are the sole bundle detected with the full bundle setting but not by the proposed approach. 
% In ELD, correlation trends are opposites (between $-0.6$ to $-0.5$ in IFW and $0.5$ to $0.6$ in FA). 
% Overall, the sub-bundle clustering detected more significant bundles than the standard full bundle analysis, for both IFW ($10$ vs $6$ in LLD; $4$ vs $1$ in ELD) and FA ($12$ vs $8$ in ELD). Only $5$ bundles over the $27$ bundles detected using sub-bundles profiles showed significant results in several clusters. 

\subsection{Group comparison}
\label{ssec:group_comparison}

\subsubsection{ELD}
\begin{table}[H]
    \centering
    \begin{tabular}{llccccc}
\toprule
Metric & Comparison & Clustering & N. sig. bundles & \% sig. pts & Mean diff. \\
\midrule
\multirow{6}{*}{FA} & \multirow{2}{*}{$\mathrm{Remission} - \mathrm{Control}$} & $\text{Full bundle}$ & 8 & 16.0\% & 2.2 $\pm$ 0.4 * \\
 &  & $\text{Sub-bundles}$ & 5 (2 multi.) & 15.4\% & 2.5 $\pm$ 0.7 * \\
\cmidrule(l){2-6}
 & \multirow{2}{*}{$\mathrm{Resistant} - \mathrm{Control}$} & $\text{Full bundle}$ & 2 & 10.0\% & -0.1 $\pm$ 3.9 * \\
 &  & $\text{Sub-bundles}$ & 1 & 8.0\% & -1.6 * \\
\cmidrule(l){2-6}
 & \multirow{2}{*}{$\mathrm{Remission} - \mathrm{Resistant}$} & $\text{Full bundle}$ & 6 & 10.8\% & 1.8 $\pm$ 0.3 * \\
 &  & $\text{Sub-bundles}$ & 11 & 8.6\% & 2.9 $\pm$ 0.6 * \\
\bottomrule
\end{tabular}

    \caption{Summary of significant bundles identified by each tractometry clustering in ELD. For each clustering, we report the number of significant bundles (N. sig. bundles)(out of 71) reaching statistical significance (AFQ-corrected point clusters with at least one point-level FWE-corrected point). For the proposed clustering, the number of bundles with at least 2 significant sub-bundles is also reported in parentheses. The percentage of significant points (\% sig. pts) is the ratio of significant to total evaluated points, averaged across bundles. The mean difference (Mean diff.) is the average point-wise group difference across all significant points (average $p < 0.05$ as * with multiple comparison).}
    \label{tab:global_stat_tracto_amynet}
\end{table}

\begin{figure}
    \centering
    \begin{minipage}[t]{0.49\textwidth}
        \vspace{0pt}
        \resizebox{\textwidth}{!}{
        \begin{tabular}{llcccc}
\toprule
 &  & \multicolumn{2}{c}{$\text{Full-bundle}$} & \multicolumn{2}{c}{$\text{Sub-bundles}$} \\
\cmidrule(lr){3-4} \cmidrule(lr){5-6}
Comparison & Bundle & $\overline{\Delta}$ & $\overline{n}_{\text{sig}}$ & $\overline{\Delta}$ & $\overline{n}_{\text{sig}}$ \\
\midrule
\multirow{9}{*}{$\mathrm{Remission - Control}$} & $\mathrm{ATR\--left}$ & $2.9^{**}_{\%}$ & $10$ & $\mathbf{3.5^{**}_{\%}}$ & $8_{\times 2}$ \\
 & $\mathrm{CC\--1}$ & $\mathbf{2.1^{*}_{\%}}$ & $9$ & -- & -- \\
 & $\mathrm{ST\--FO\--left}$ & $2.3^{**}_{\%}$ & $32$ & $\mathbf{2.4^{*}_{\%}}$ & $38$ \\
 & $\mathrm{ST\--FO\--right}$ & $\mathbf{2.6^{*}_{\%}}$ & $9$ & -- & -- \\
 & $\mathrm{ST\--PREF\--left}$ & $1.7^{*}_{\%}$ & $33$ & $\mathbf{1.9^{*}_{\%}}$ & $16_{\times 2}$ \\
 & $\mathrm{ST\--PREM\--left}$ & $1.7^{**}_{\%}$ & $9$ & $\mathbf{3.0^{**}_{\%}}$ & $9$ \\
 & $\mathrm{T\--PREF\--left}$ & $\mathbf{2.1^{*}_{\%}}$ & $13$ & -- & -- \\
 & $\mathrm{UF\--left}$ & $\mathbf{2.0^{*}_{\%}}$ & $13$ & -- & -- \\
 & $\mathrm{CC\--7}$ & -- & -- & $\mathbf{1.8^{*}_{\%}}$ & $6$ \\
\midrule
\multirow{3}{*}{$\mathrm{Resistant - Control}$} & $\mathrm{OR\--left}$ & $\mathbf{-2.9^{*}_{\%}}$ & $10$ & -- & -- \\
 & $\mathrm{ST\--FO\--right}$ & $\mathbf{2.6^{*}_{\%}}$ & $10$ & -- & -- \\
 & $\mathrm{ST\--POSTC\--left}$ & -- & -- & $\mathbf{-1.6^{*}_{\%}}$ & $8$ \\
\midrule
\multirow{11}{*}{$\mathrm{Remission - Resistant}$} & $\mathrm{CC\--2}$ & $1.8^{*}_{\%}$ & $11$ & $\mathbf{3.0^{**}_{\%}}$ & $9$ \\
 & $\mathrm{CC\--3}$ & $1.9^{*}_{\%}$ & $8$ & $\mathbf{3.2^{*}_{\%}}$ & $6$ \\
 & $\mathrm{CG\--left}$ & $1.7^{*}_{\%}$ & $15$ & $\mathbf{3.4^{*}_{\%}}$ & $8$ \\
 & $\mathrm{CG\--right}$ & $2.3^{**}_{\%}$ & $10$ & $\mathbf{4.0^{**}_{\%}}$ & $8$ \\
 & $\mathrm{ST\--PREF\--left}$ & $1.6^{**}_{\%}$ & $11$ & $\mathbf{2.2^{**}_{\%}}$ & $17$ \\
 & $\mathrm{ST\--PREM\--left}$ & $1.6^{*}_{\%}$ & $10$ & $\mathbf{2.3^{*}_{\%}}$ & $9$ \\
 & $\mathrm{CC\--4}$ & -- & -- & $\mathbf{2.9^{*}_{\%}}$ & $6$ \\
 & $\mathrm{CC\--5}$ & -- & -- & $\mathbf{2.2^{*}_{\%}}$ & $5$ \\
 & $\mathrm{FX\--left}$ & -- & -- & $\mathbf{3.2^{*}_{\%}}$ & $9$ \\
 & $\mathrm{ICP\--left}$ & -- & -- & $\mathbf{3.2^{*}_{\%}}$ & $7$ \\
 & $\mathrm{ST\--POSTC\--left}$ & -- & -- & $\mathbf{2.2^{**}_{\%}}$ & $11$ \\
\bottomrule
\end{tabular}

        }
    \end{minipage}
    \hfill
    \begin{minipage}[t]{0.49\textwidth}
        \vspace{0pt}
        \includegraphics[width=\textwidth]{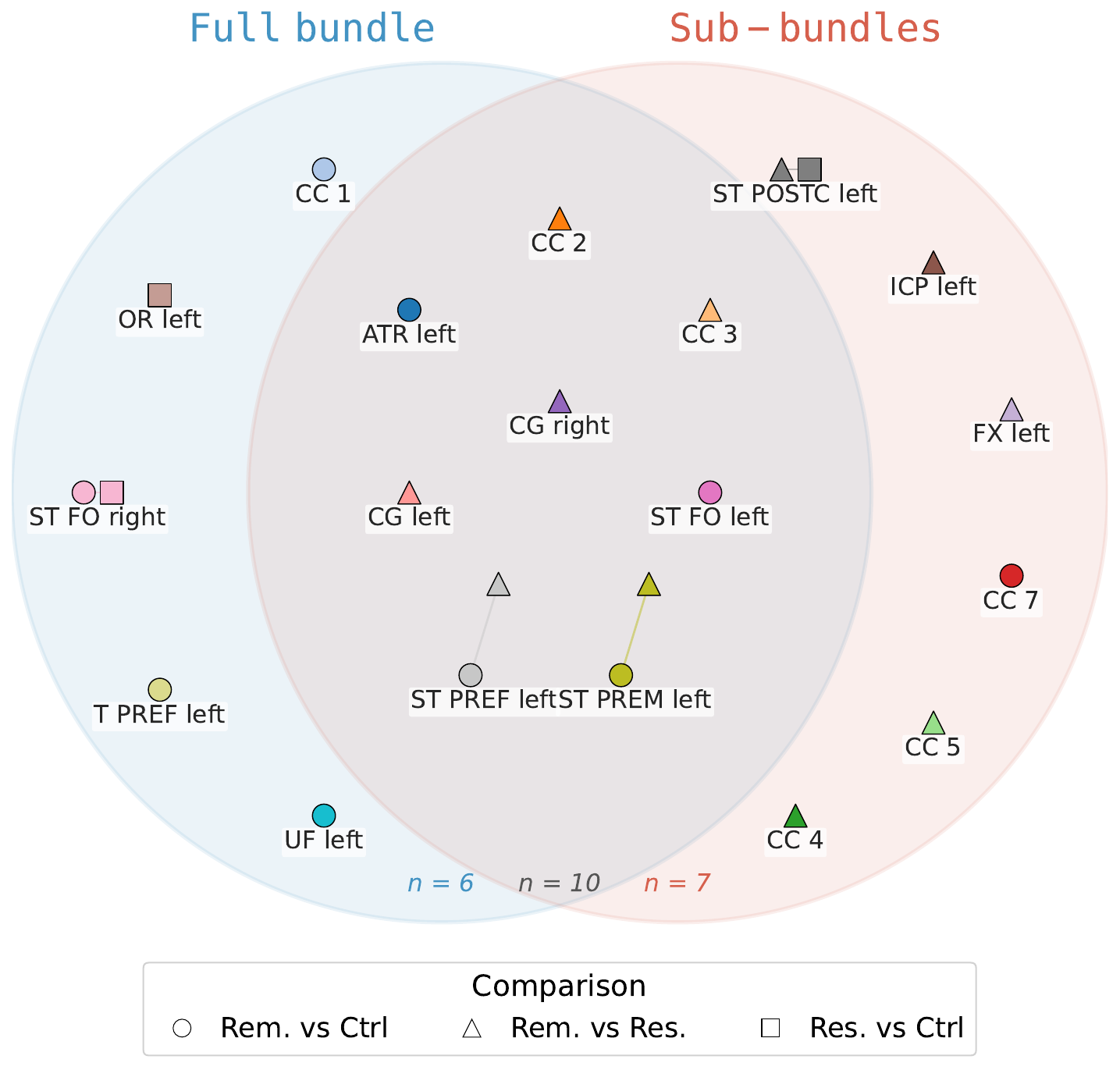}
    \end{minipage}
    \caption{(\textit{left}) Comparison of significant bundles identified by the sub-bundle and full bundle approaches for FA in ELD. Compared statistics are the average group difference in significant point ($\overline{\Delta}$) and mean significant cluster size ($\overline{n}_{\text{sig}}$). When applicable, the number of significant cluster per bundle is indicated as a multiple of the mean cluster size. (\textit{right}) Venn diagram showing the overlap of significant bundles. The shape of the displayed points distinguishes between pairwise comparisons between groups : remission vs. controls, resistant vs. controls, remission vs. resistant.}
    \label{fig:bundle_diff_amynet}
\end{figure}

The global point-wise group comparison analysis (Table \ref{tab:global_stat_tracto_amynet}) revealed that the proposed cluster approach identified a similar number of significant bundles to the full bundle analysis (17 vs 16 bundles for FA). However, it detected larger average microstructural differences, with absolute FA percentage differences increasing from 1.6\% to 2.9\% for the sub-bundle profiles and from 0.1\% to 2.2\% for the full bundle analysis. This effect was particularly pronounced in the Resistant vs.~Remission comparison, where the average difference increased from $1.8\%\pm0.3$ to $2.9\%\pm0.6$.
% , independant $t$-test $p<0.001$). 
% Only one bundle shows a significant difference between Resistant and Controls, with a decrease of FA of $-1.56\%$ for $\text{Fréchet}$, while two bundles show significant differences for Centerline but with opposite effect directions (mean effect of $-0.15\%$).

The Venn diagram (Fig.~\ref{fig:bundle_diff_amynet}) further shows that the sub-bundle analysis identified a largely overlapping set of significant bundles compared to the full bundle approach, while also revealing method-specific findings (6 bundles unique to full bundle analysis and 7 unique to clustered approach). Notably, the clustering method detected substantially more significant bundles in the Resistant vs.~Remission comparison (10 vs.~5), suggesting increased sensitivity to localized microstructural alterations.

Fig.~\ref{fig:along_tract_amynet_FA} shows along-tract profiles of significant bundle clusters in FA. It reveals FA alterations concentrated in sub-bundles connecting striatal (ST, 5 bundles), prefrontal (PREF, 2 bundles) regions. In the remission group relative to controls, FA increases are observed in striato-frontal connections, including ST-FO (left and right), ST-PREF-left, and ST-PREM-left, suggesting restored or preserved white matter integrity in remitting patients. Notably, ST-PREF-left and ST-POSTC-left each appear as significant in two distinct comparisons: ST-PREF-left is identified in both the remission vs. controls and the remission vs. resistant comparisons, while ST-POSTC-left emerges in both the resistant vs. controls and the remission vs. resistant settings.

\subsubsection{LLD}

\begin{table}[H]
    \centering
    \begin{tabular}{llccccc}
\toprule
Comparison & Metric & Clustering & N. sig. bundles & \% sig. pts & Mean diff. \\
\midrule
\multirow{4}{*}{LLD - Control} & \multirow{2}{*}{FA} & Full bundle & 4 & 12.5\% & -2.2 $\pm$ 0.6 * \\
 &  & Sub-bundles & 6 & 14.8\% & -2.4 $\pm$ 0.9 * \\
\cmidrule(l){2-6}
 & \multirow{2}{*}{IFW} & Full bundle & 15 & 13.6\% & 2.1 $\pm$ 0.5 * \\
 &  & Sub-bundles & 23 (5 multi.) & 13.0\% & 2.5 $\pm$ 0.8 * \\
\bottomrule
\end{tabular}

    \caption{Summary of significant bundles identified by each tractometry clustering in LLD. For each clustering, we report the number of bundles (out of 71) reaching statistical significance (AFQ-corrected point clusters with at least one point-level FWE-corrected point). For the proposed clustering, the number of bundles with at least 2 significant sub-bundles is also reported in parentheses. The percentage of significant points is the ratio of significant to total evaluated points, averaged across bundles. The mean difference is the average point-wise group difference across all significant points (average $p < 0.05$ as * with multiple comparison).}
    \label{tab:global_stat_tracto_actidep}
\end{table}

\begin{figure}
    \centering
    \begin{subfigure}{\textwidth}
        \centering
        \begin{minipage}[t]{0.44\textwidth}
            \vspace{60pt}
            \resizebox{\textwidth}{!}{
            \begin{tabular}{llcccc}
\toprule
 &  & \multicolumn{2}{c}{$\text{Full-bundle}$} & \multicolumn{2}{c}{$\text{Sub-bundles}$} \\
\cmidrule(lr){3-4} \cmidrule(lr){5-6}
Comparison & Bundle & $\overline{\Delta}$ & $\overline{n}_{\text{sig}}$ & $\overline{\Delta}$ & $\overline{n}_{\text{sig}}$ \\
\midrule
\multirow{6}{*}{$\mathrm{LLD - Control}$} & $\mathrm{CC\--1}$ & $-2.9^{*}_{\%}$ & $8$ & $\mathbf{-4.1^{**}_{\%}}$ & $8$ \\
 & $\mathrm{MLF\--left}$ & $-1.4^{*}_{\%}$ & $7$ & $\mathbf{-1.7^{*}_{\%}}$ & $10$ \\
 & $\mathrm{ST\--FO\--left}$ & $-2.3^{**}_{\%}$ & $23$ & $\mathbf{-2.6^{*}_{\%}}$ & $16$ \\
 & $\mathrm{ST\--FO\--right}$ & $-2.2^{*}_{\%}$ & $12$ & $\mathbf{-2.5^{*}_{\%}}$ & $17$ \\
 & $\mathrm{ST\--PAR\--right}$ & -- & -- & $\mathbf{-1.6^{*}_{\%}}$ & $10$ \\
 & $\mathrm{ST\--PREF\--left}$ & -- & -- & $\mathbf{-1.9^{*}_{\%}}$ & $28$ \\
\bottomrule
\end{tabular}

            }
        \end{minipage}
        \hfill
        \begin{minipage}[t]{0.55\textwidth}
            \vspace{0pt}
            \includegraphics[width=\textwidth]{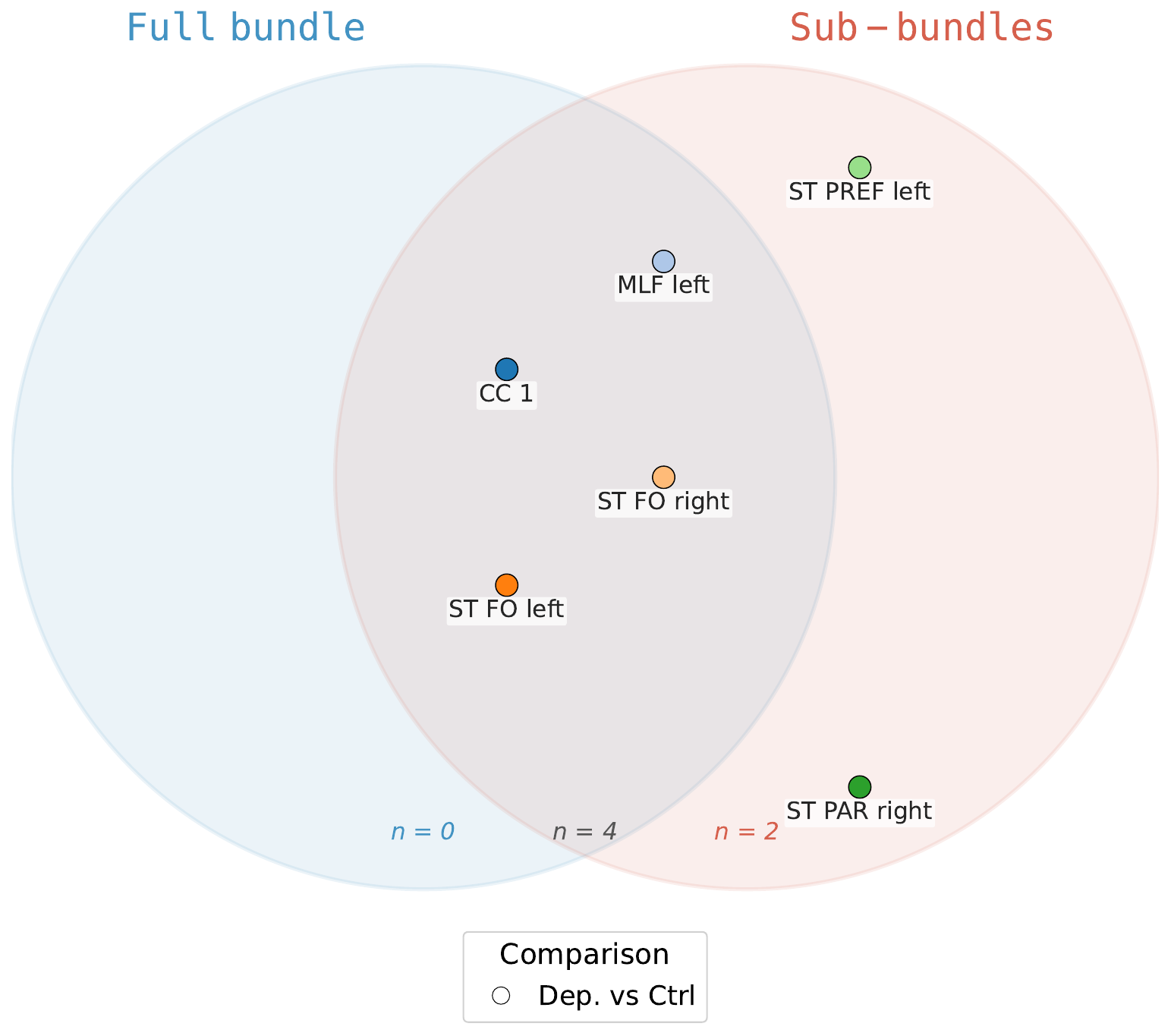}
        \end{minipage}
        \caption{FA}
    \end{subfigure}
    \vskip\baselineskip
    \begin{subfigure}{\textwidth}
        \centering
        \begin{minipage}{0.44\textwidth}
            \vspace{0pt}
            \resizebox{\textwidth}{!}{
            \begin{tabular}{llcccc}
\toprule
 &  & \multicolumn{2}{c}{$\text{Full-bundle}$} & \multicolumn{2}{c}{$\text{Sub-bundles}$} \\
\cmidrule(lr){3-4} \cmidrule(lr){5-6}
Comparison & Bundle & $\overline{\Delta}$ & $\overline{n}_{\text{sig}}$ & $\overline{\Delta}$ & $\overline{n}_{\text{sig}}$ \\
\midrule
\multirow{24}{*}{$\mathrm{LLD - Control}$} & $\mathrm{CC\--1}$ & $\mathbf{2.8^{*}_{\%}}$ & $13$ & $2.4^{*}_{\%}$ & $9.5_{\times 2}$ \\
 & $\mathrm{CC\--3}$ & $2.0^{*}_{\%}$ & $12$ & $\mathbf{2.0^{*}_{\%}}$ & $9$ \\
 & $\mathrm{IFO\--right}$ & $2.2^{*}_{\%}$ & $9$ & $\mathbf{2.9^{*}_{\%}}$ & $8$ \\
 & $\mathrm{POPT\--right}$ & $\mathbf{2.8^{*}_{\%}}$ & $13$ & $2.7^{*}_{\%}$ & $8$ \\
 & $\mathrm{SLF\--III\--right}$ & $2.1^{*}_{\%}$ & $11$ & $\mathbf{2.5^{**}_{\%}}$ & $12$ \\
 & $\mathrm{ST\--FO\--left}$ & $\mathbf{1.7^{*}_{\%}}$ & $11$ & -- & -- \\
 & $\mathrm{ST\--PAR\--right}$ & $1.7^{*}_{\%}$ & $14$ & $\mathbf{1.9^{*}_{\%}}$ & $15.5_{\times 2}$ \\
 & $\mathrm{ST\--POSTC\--right}$ & $1.5^{**}_{\%}$ & $14$ & $\mathbf{1.5^{*}_{\%}}$ & $12_{\times 2}$ \\
 & $\mathrm{ST\--PREC\--right}$ & $1.6^{*}_{\%}$ & $14$ & $\mathbf{1.8^{**}_{\%}}$ & $16$ \\
 & $\mathrm{ST\--PREF\--left}$ & $1.5^{**}_{\%}$ & $26$ & $\mathbf{1.6^{*}_{\%}}$ & $14.5_{\times 2}$ \\
 & $\mathrm{STR\--right}$ & $2.1^{**}_{\%}$ & $16$ & $\mathbf{2.2^{*}_{\%}}$ & $19$ \\
 & $\mathrm{T\--PAR\--right}$ & $2.4^{**}_{\%}$ & $14$ & $\mathbf{3.4^{**}_{\%}}$ & $19_{\times 2}$ \\
 & $\mathrm{T\--POSTC\--right}$ & $1.5^{*}_{\%}$ & $10$ & $\mathbf{1.6^{*}_{\%}}$ & $12$ \\
 & $\mathrm{T\--PREC\--right}$ & $1.8^{*}_{\%}$ & $12$ & $\mathbf{1.9^{*}_{\%}}$ & $17$ \\
 & $\mathrm{T\--PREM\--right}$ & $3.1^{*}_{\%}$ & $15$ & $\mathbf{3.3^{*}_{\%}}$ & $14$ \\
 & $\mathrm{CST\--right}$ & -- & -- & $\mathbf{2.1^{**}_{\%}}$ & $10$ \\
 & $\mathrm{FPT\--left}$ & -- & -- & $\mathbf{4.5^{**}_{\%}}$ & $7$ \\
 & $\mathrm{OR\--right}$ & -- & -- & $\mathbf{3.8^{*}_{\%}}$ & $22$ \\
 & $\mathrm{ST\--PREF\--right}$ & -- & -- & $\mathbf{2.0^{*}_{\%}}$ & $16$ \\
 & $\mathrm{ST\--PREM\--left}$ & -- & -- & $\mathbf{2.5^{*}_{\%}}$ & $9$ \\
 & $\mathrm{T\--PREF\--left}$ & -- & -- & $\mathbf{2.0^{*}_{\%}}$ & $16$ \\
 & $\mathrm{T\--PREF\--right}$ & -- & -- & $\mathbf{2.6^{**}_{\%}}$ & $11$ \\
 & $\mathrm{T\--PREM\--left}$ & -- & -- & $\mathbf{2.9^{*}_{\%}}$ & $11$ \\
 & $\mathrm{UF\--left}$ & -- & -- & $\mathbf{2.9^{*}_{\%}}$ & $12$ \\
\bottomrule
\end{tabular}

            }
        \end{minipage}
        \hfill
        \begin{minipage}{0.55\textwidth}
            \vspace{0pt}
            \includegraphics[width=\textwidth]{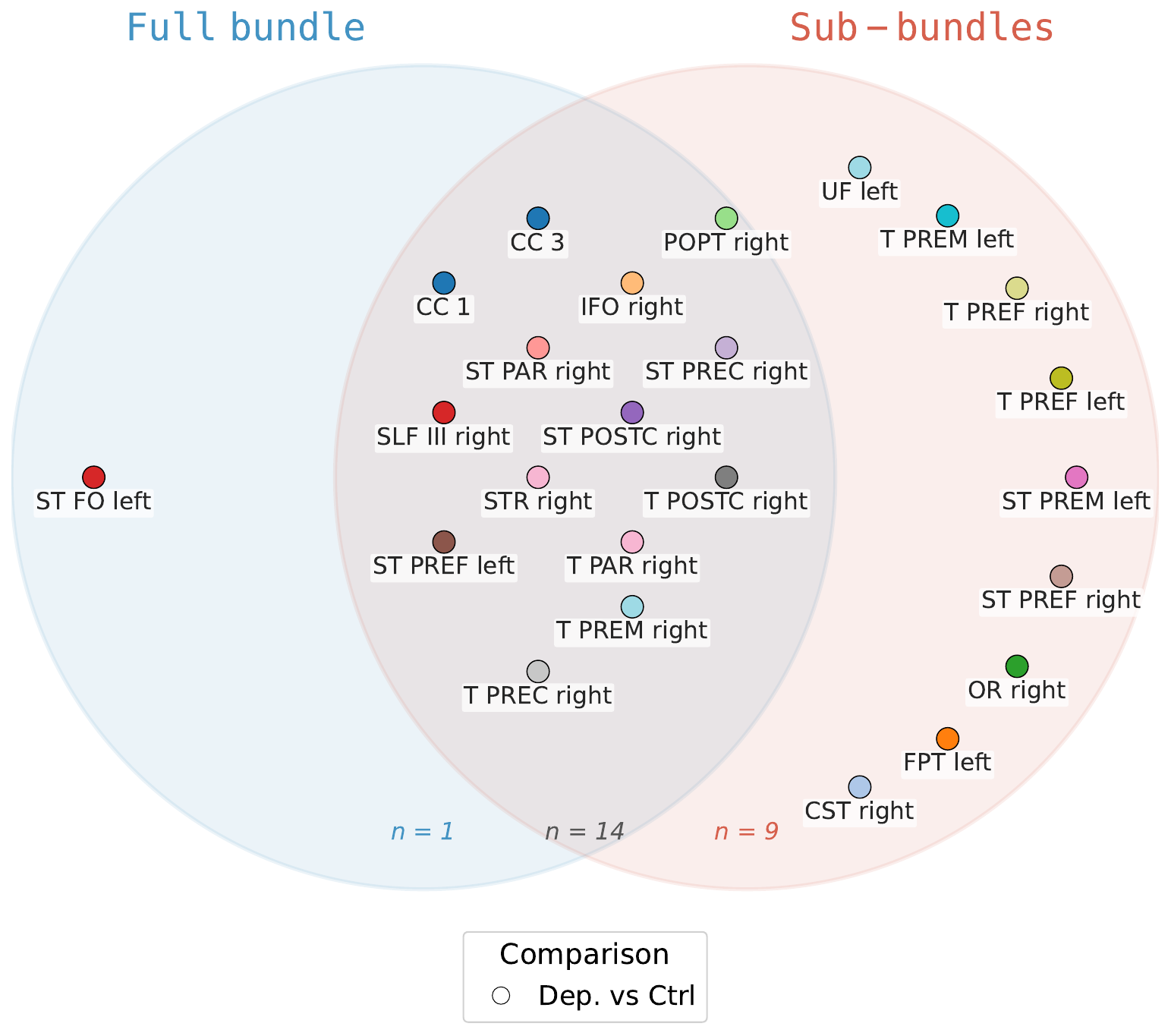}
        \end{minipage}
        \caption{IFW}
    \end{subfigure}
    \caption{Comparison of significant bundles identified by the sub-bundle and full bundle approaches in LLD (LLD vs. controls). Left: per-bundle along-tract summary table. Compared statistics are the average group difference in significant point ($\overline{\Delta}$) and mean significant cluster size ($\overline{n}_{\text{sig}}$). When applicable, the number of significant cluster per bundle is indicated as a multiple of the mean cluster size. Right: Venn diagram showing the overlap of significant bundles between approaches. Excluded bundles for analysis : left and right ILF, left and right FX, CA, left SLF I, right SLF II, and left SLF III.  (a)~FA; (b)~IFW.}
    \label{fig:bundle_diff_actidep}
\end{figure}

The global results of the group comparison analysis (Table ~\ref{tab:global_stat_tracto_actidep}) revealed that using sub-bundles increase the number of significant results in distinct bundles for both FA (6 vs. 4) and IFW (23 vs. 15), with a higher mean difference in both metrics (FA: $-2.4\%$ vs. $-2.2\%$; IFW: $2.5\%$ vs. $2.1\%$). The Venn diagram analysis (Fig.~\ref{fig:bundle_diff_actidep}) shows that the sub-bundle clustering method identifies a superset of those found by the full bundle, with additional bundles (2 in FA, 9 in IFW). Only the ST FO left bundle (out of $23$ bundles) detected by the standard analysis is not identified in sub-bundles, and only for IFW. Along-tract profiles visualization are available in the supplementary material (Fig.~\ref{fig:along_tract_actidep_IFW} and ~\ref{fig:along_tract_actidep_FA}).

Figure~\ref{fig:along_tract_actidep_IFW} presents the along-tract profiles of significant bundle clusters for IFW. The results reveal a marked increase in IFW in the LLD group relative to controls, localized to specific sub-bundles connecting distinct brain regions, including the striatum (ST; 6 of 23 significant bundles), thalamus (T; 7 bundles), prefrontal cortex (PREF; 4 bundles), and premotor cortex (PREM; 3 bundles).

For FA, as illustrated in Fig.~\ref{fig:along_tract_actidep_FA}, cluster ~1 of the left ST–PREF and right ST–PAR pathways identified localized FA reductions in the LLD group that were not detected by the full bundle approach. 

These findings demonstrate that the clustering framework can reveal localized microstructural alterations that are not detected by the full bundle approach.

\section{Discussion}

%This work introduces a bundle clustering tractometry framework that employs a hierarchical clustering with an automatic, quality-driven determination of the number of sub-bundles, eliminating the need for empirical geometric thresholds. Separation is based on the Fréchet distance, which measures maximum trajectory deviations rather than average distances, enabling the identification of well-separated sub-bundle groups.

This work introduces a bundle-clustering tractometry framework based on hierarchical clustering with an automatic, quality-driven determination of the number of sub-bundles, thereby eliminating the need for empirically defined geometric thresholds. Clustering is performed using the Fréchet distance, which captures the maximum deviation between streamline trajectories rather than their average separation. This approach enables the identification of well-separated and anatomically distinct sub-bundle groups.

% To validate this approach, we study the interest of this approach by analyzing the microstructural homogeneity, age sensitivity and the microstructural differences within cluster. Second, we evaluate its ability to capture microstructural white matter alterations in two depression cohorts spanning different age ranges and clinical contexts: ELD, a treatment-resistance and remitted cohort of middle-aged adults, and LLD, a cohort of elderly patients with late-life depression. 

% The results demonstrate that the proposed approach offers advantages over conventional whole bundle centerline methods, with benefits that depend on the evaluated microstructural metric and the cohort. 

\paragraph{Microstructural metrics are not homogeneous across sub-bundles.}

A key motivation for the sub-bundle approach is the hypothesis that microstructural alterations are spatially heterogeneous within white matter bundles and may not be uniformly distributed along the streamline trajectories. Analysis of the microstructural SD within sub-bundles compared with the corresponding full bundles (Fig.~\ref{fig:std_best_centroid}) confirms that clustering generally produces more homogeneous sub-bundles. These results suggest that reducing a bundle to a single representative profile may obscure meaningful intra-bundle spatial variability, whereas the proposed clustering strategy preserves and characterizes this heterogeneity across diverse populations and microstructural metrics. 

\paragraph{Sub-bundle analysis is more sensitive to age-related microstructural changes.}
% Both cohorts exhibit age-related microstructural variations, but with opposite directions consistent with their respective age ranges: in LLD, age correlates positively with IFW and negatively with FA, matching the canonical aging pattern, whereas in ELD the trends are reversed, which may reflects late white matter maturation in younger adults~\citep{kochunov2010fractional}. 
% In both cases, the sub-bundle analysis preserves the direction reported by the whole-bundle approach, indicating no apparent false positives. Crucially, it also detects additional bundles, and these new detections systematically arise from specific clusters rather than spread across the entire bundle, confirming that age-related effects are spatially heterogeneous and emerge at the sub-bundle scale.

Both microstructural metrics exhibit age-related variations, but with opposite directions, i.e an increase in IFW and a decrease in FA with age. This is consistent with their biological interpretation, namely the neuroinflammation for IFW and the loss of fiber organization and integrity for FA, and is supported by the litterature~\citep{kochunov2010fractional}. In both cases, the sub-bundle analysis preserves the direction reported by the whole-bundle approach, indicating no apparent false positives. Crucially, it also detects additional bundles, and these new detections systematically arise from specific clusters rather than spread across the entire bundle, confirming that age-related effects are spatially heterogeneous and emerge at the sub-bundle scale.

\paragraph{Sub-bundle analysis reveals localized group differences in fanning topologies}

The gain of the sub-bundle approach over the full bundle profile analysis is more pronounced in LLD. This cohort-dependent gap is methodologically informative. By construction, the proposed clustering partitions streamlines by trajectory divergence, making it particularly sensitive to fanning topographies, in which streamlines progressively diverge towards spatially distributed cortical terminations. Such configurations are typical of thalamo-cortical and striato-cortical projections, and of association fibers. Some of these bundles had previously been identified in this cohort in a study by \citet{hedouin2024microstructural}, but the proposed method enabled additional significant differences in other bundles in this region to be detected.
Consistent with this rationale, the sub-bundle-specific findings in LLD are predominantly localized within fanning fiber configurations. Thalamo-cortical and striato-cortical projections (T-PREF, T-PREM, ST-PREF, ST-PREM, ST-PAR) account for most of the IFW bundles detected exclusively  by the sub-bundle approach (Fig.~\ref{fig:bundle_diff_actidep}). This spatial pattern matches with the concept of differential brain aging~\citep{raz2006differential}, which describes non-uniform vulnerability of white matter, particularly affecting frontal and superficial regions. This structural pattern also aligns with our previous functional findings, which demonstrated that motor activity, a marker of depression severity, in LLD were primarily associated with altered resting-state functional connectivity within the prefrontal, premotor, and parietal cortices of the default mode, cingulo-opercular, and frontoparietal networks~\citep{roy2024quantifying}. Free-water content in these regions is also strongly sensitive to partial volume effects with cortical grey matter.

In ELD, the alterations appear concentrated in deep white matter regions, notably in the corpus callosum subdivisions (CC1-CC7). These regions are dominated by parallel, geometrically coherent fibers rather than fanning, which limits the advantage of trajectory-divergence-based clustering. 

Furthermore, only FA differences were detected in these younger patients, whereas the majority of differences in LLD were observed in IFW. These results align with the fact that IFW is associated with neuroinflammation~\citep{sumra2025regional}, a mechanism closely linked to brain aging that is more pronounced in individuals susceptible to LLD~\citep{vu2013depression}. Importantly, the absence of IFW findings in ELD when using our proposed approach demonstrates its selectivity, indicating that the method does not detect arbitrary changes.

\paragraph{Limitations and perspectives.}

This study has several limitations. First, the proposed methodology for deriving subject-specific bundle clusters depends on both the quality of the atlas and the accuracy of the registration between individual subjects and the atlas space. Registration inaccuracies may introduce systematic errors in streamline assignment and cluster correspondence, thereby affecting the reliability of the resulting microstructural profiles. The development of registration methods specifically tailored to tractography data therefore remains an important challenge for improving the robustness and reproducibility of atlas-based tractometry analyses.

A related limitation concerns the treatment of sub-bundles that could not be assigned across all subjects. To ensure a robust and homogeneous basis for statistical comparison, these incomplete sub-bundles were excluded from further analysis. Although this strategy avoids unbalanced sample sizes across clusters, it may introduce selection bias by removing potentially informative sub-bundles. The origin of these unassigned clusters warrants further investigation to determine whether they arise from segmentation inaccuracies, tractography failures, registration errors, or genuine inter-individual anatomical variability. Bundles that had at least one missing subject per sub-bundle are illustrated in the supplementary material (Fig.~\ref{fig:excluded_bundles}). Bundles such as the CA and bilateral FX exhibit largely overlapping sub-bundles separated only by fiber length variations, indicating that clustering is simply inappropriate for these tracts. Conversely, for larger bundles (ILF, left SLF I, right SLF II, left SLF III), missing associations are likely due to tractography failures. This is supported by the issue being exclusive to the elderly cohort, where neuroinflammation may degrade the diffusion signal.

In the present study, all 71 bundles were systematically subdivided to enable a rigorous and unbiased comparison between sub-bundle and full bundle tractometry. However, this design choice may not be optimal for all white matter pathways. Smaller and anatomically compact tracts, such as the CA or the FX, may exhibit limited internal spatial heterogeneity and therefore derive less benefit from further subdivision. Future work should investigate adaptive clustering strategies in which the number of sub-bundles is determined using data-driven model selection criteria, allowing the level of subdivision to better reflect the intrinsic anatomical and geometric complexity of each bundle. Moreover, the current evaluation is restricted to FA and IFW; other microstructural metrics (e.g. neurite density) may show different patterns of sensitivity across approaches and populations. 

The statistical robustness of the reported findings deserves consideration. The statistical robustness of the reported findings deserves consideration. Because our inference strategy relies on a conservative clusterFWE correction applied independently within each bundle or sub-bundle profile, the local risk of Type I errors is equally controlled in both the full bundle and sub-bundle approaches. Consequently, the increased number of significant bundles detected by the sub-bundle method is unlikely to be driven by an inflation of false positives, but rather reflects a genuine gain in sensitivity due to reduced signal averaging. However, in the sub-bundle approach, a bundle is declared significant if at least one of its sub-bundles yields a significant cluster, which slightly inflates the family-wise error rate at the bundle level. Extending the multiple comparison correction (e.g., the AFQ cluster-based framework) across clusters within the same bundle, as suggested by Zheng et al.~\citep{zheng2026agfstractometry}, would be a natural avenue to rigorously control this residual this risk, but would require a prior characterisation of spatial neighbourhoods between clusters of the same bundle.

% We hypothesise that the advantage of $\text{Fréchet}$-based clustering relies primarily from its ability to segment bundles exhibiting fanning topographies, where the single-centerline assumption of homogeneous cross-sectional microstructure is most strongly violated. If this hypothesis holds, a natural extension would be a hybrid clustering scheme that combines the geometric strengths of both approaches: a first stage separating fibres with spatially distinct $\text{Endpoint}$s (capturing crossing topographies), followed by a second stage imposing a maximum within-cluster $\text{Fréchet}$ width on each resulting sub-group (capturing fanning topographies). Such a two-stage strategy would explicitly address the two principal sources of intra-bundle geometric heterogeneity within a unified framework, and constitutes a promising direction for future work.

Finally, although the clustering approach increases the spatial resolution of the analysis compared to a full bundle profile analysis, the framework remains fundamentally point-wise in nature, aggregating microstructural measurements at discrete locations along the tract. More integrative approaches exploiting the full three-dimensional geometry of a bundle could provide a richer characterization of white matter organization and may better capture alterations that are distributed across the spatial extent of a bundle rather than localized along its longitudinal axis.

\section{Conclusion}
We have presented a sub-bundle tractometry framework that decomposes each bundle into geometrically coherent sub-bundles. The decomposition is fully automatic:  the optimal number of sub-bundles is selected by maximizing cluster separation, without requiring any per-bundle parameter tuning.

Validation on two independent depression cohorts confirms that sub-bundles produces more homogeneous clusters than within the full bundle in FA and IFW. The sub-bundle approach highlights more bundles with significant age correlations in both cohorts and both metrics. New detections arise from specific sub-bundles rather than from the whole tract, showing that age-related effects are spatially heterogeneous and can be missed when the entire bundle is summarised by a single profile. The gain in group discriminability is larger in the late-life depression cohort, where aging-related changes affect fanning projection tracts that proposed bundle clustering is designed to resolve. It is more modest in the younger treatment-resistance cohort, where alterations concentrate in deep, geometrically coherent tracts such as the corpus callosum.

These results demonstrate that performing tractometry at a finer spatial scale than the whole bundle is beneficial, and that automatic clustering is a practical way to achieve this in group studies where white matter alterations are subtle and spatially heterogeneous, such as in diffusion measure in aged participants.

\section*{Author contributions statement}
N.D. performed the statistics, interpreted the results and wrote the first draft; J-C.R. and T.D. designed the study, performed the data acquisition and revised the manuscript; M-L.P.M found the financial supports and revised the manuscript;  G.H.R. and J.C designed the study, found the financial supports, authorization, data acquisition, supervised this study, interpreted the results and revised the manuscript.

\section*{Ethics}
All methods were performed in accordance with the guidelines of the Declaration of Helsinki, and the study was approved by the relevant institutional review boards (ID-RCB 2018-AO2643-52, NCT03807167 and IDR-RCB 2013-A00662-43). All subjects provided informed, written consent before participation in the study, according to the French legislation.

\section*{Data and Code Availability}
The original datasets used and analyzed as part of this study are available from the corresponding author, subject to approval by the local ethics committee of the researcher making the request. Aggregated data and the code used to generate the figures in this paper are publicly available at \url{https://github.com/nathandecaux/AutoSubTracto}. The code used to process the original datasets is available at \url{https://github.com/nathandecaux/TractoPL}. 

\section*{Funding}

This study was supported by the Fondation de l'Avenir, Fondation Planiol, Fondation de France, and the Comité de recherche Clinique et d'Evaluation Thérapeutique from the Rennes University Hospital. J-CR and ND received financial support from the Institut de Neuroscience Clinique de Rennes. 

\section*{Declaration of Competing interests}
The authors declare no competing interests.

\section*{Declaration of the use of AI}

We did not use AI in writing this paper.
 
\printbibliography

@article{vu2013depression,
   author = {Nicholas Q. Vu and Howard J. Aizenstein},
   doi = {10.1097/WCO.0000000000000028},
   issn = {13507540},
   issue = {6},
   journal = {Current Opinion in Neurology},
   month = {12},
   pages = {656-661},
   pmid = {24184971},
   title = {Depression in the elderly: Brain correlates, neuropsychological findings, and role of vascular lesion load},
   volume = {26},
   year = {2013}
}

@article{joo2025alongtract,
   author = {Sung Woo Joo and Hyeongyu Park and Jihyu Park and Jungsun Lee},
   doi = {10.1038/s41537-025-00586-1},
   issn = {2754-6993},
   issue = {1},
   journal = {Schizophrenia 2025 11:1},
   month = {3},
   pages = {37-},
   publisher = {Nature Publishing Group},
   title = {Along-tract white matter abnormalities and their clinical associations in recent-onset and chronic schizophrenia},
   volume = {11},
   url = {https://www.nature.com/articles/s41537-025-00586-1},
   year = {2025}
}

@article{tones2005pasta,
   author = {Derek K. Tones and Adam R. Travis and Greg Eden and Carlo Pierpaoli and Peter J. Basser},
   doi = {10.1002/MRM.20484},
   issn = {0740-3194},
   issue = {6},
   journal = {Magnetic resonance in medicine},
   pages = {1462-1467},
   pmid = {15906294},
   publisher = {Magn Reson Med},
   title = {PASTA: pointwise assessment of streamline tractography attributes},
   volume = {53},
   url = {https://pubmed.ncbi.nlm.nih.gov/15906294/},
   year = {2005}
}

@article{chang2025free,
   author = {Kelly Chang and Luke Burke and Nina LaPiana and Bradley Howlett and David Hunt and Margaret Dezelar and Jalal B. Andre and Patti Curl and James D. Ralston and Ariel Rokem and Christine L. Mac Donald},
   doi = {10.1162/IMAG.A.991/133658/FREE-WATER-ELIMINATION-TRACTOMETRY-FOR-AGING},
   issn = {28376056},
   journal = {Imaging Neuroscience},
   month = {11},
   publisher = {Massachusetts Institute of Technology},
   title = {Free water elimination tractometry for aging brains},
   volume = {3},
   url = {https://dx.doi.org/10.1162/IMAG.a.991},
   year = {2025}
}

@article{witt2025tractspecific,
   author = {Atlee A. Witt and Sawyer Fleishman and Delaney Houston and Logan E. Prock and Grace Sweeney and Trey McGonigle and Simon Vandekar and Maxime Chamberland and Seth Stubblefield and Colin D. McKnight and Kristin P. O’Grady and Kurt Schilling and Seth A. Smith},
   doi = {10.1162/IMAG.A.72/131446/TRACT-SPECIFIC-ANALYSIS-OF-DIFFUSION-MRI-AT-3T},
   issn = {28376056},
   journal = {Imaging Neuroscience},
   month = {7},
   pages = {2025},
   publisher = {Massachusetts Institute of Technology},
   title = {Tract-specific analysis of diffusion MRI at 3T detects cervical spinal cord aberrations in multiple sclerosis},
   volume = {3},
   url = {https://dx.doi.org/10.1162/IMAG.a.72},
   year = {2025}
}

@article{cox2016ageing,
   author = {Simon R. Cox and Stuart J. Ritchie and Elliot M. Tucker-Drob and David C. Liewald and Saskia P. Hagenaars and Gail Davies and Joanna M. Wardlaw and Catharine R. Gale and Mark E. Bastin and Ian J. Deary},
   doi = {10.1038/ncomms13629},
   issn = {2041-1723},
   issue = {1},
   journal = {Nature Communications 2016 7:1},
   month = {12},
   pages = {13629-},
   pmid = {27976682},
   publisher = {Nature Publishing Group},
   title = {Ageing and brain white matter structure in 3,513 UK Biobank participants},
   volume = {7},
   url = {https://www.nature.com/articles/ncomms13629},
   year = {2016}
}

@article{vandeloo2022freewater,
   author = {Katie L. Vandeloo and Patricia Burhunduli and Sylvain Bouix and Kimia Owsia and Kang Ik K. Cho and Zhuo Fang and Amanda Van Geel and Ofer Pasternak and Pierre Blier and Jennifer L. Phillips},
   doi = {10.1016/J.BPSC.2022.12.007},
   issn = {24519030},
   issue = {4},
   journal = {Biological psychiatry. Cognitive neuroscience and neuroimaging},
   month = {4},
   pages = {471},
   pmid = {36906445},
   publisher = {Elsevier Inc.},
   title = {Free-Water Diffusion Magnetic Resonance Imaging Differentiates Suicidal Ideation From Suicide Attempt in Treatment-Resistant Depression},
   volume = {8},
   url = {https://pmc.ncbi.nlm.nih.gov/articles/PMC11421579/},
   year = {2022}
}

@article{cao2025association,
   author = {Yuan Cao and Paulo Lizano and Meng Li and Nils Opel and Zümrüt Duygu Sen and Lejla Colic and Huan Sun and Xiaoqin Zhou and Merita Aruci and Tara Chand and Xipeng Long and Gaoju Deng and Jingshi Mu and Shuo Guo and Huaiqiang Sun and Qiyong Gong and Changjian Qiu and Martin Walter and Zhiyun Jia},
   doi = {10.1016/J.BBI.2025.04.005},
   issn = {1090-2139},
   journal = {Brain, behavior, and immunity},
   month = {8},
   pages = {208-218},
   pmid = {40199429},
   publisher = {Brain Behav Immun},
   title = {Association between peripheral inflammation and body mass index on white matter integrity and free water in bipolar II depression},
   volume = {128},
   url = {https://pubmed.ncbi.nlm.nih.gov/40199429/},
   year = {2025}
}

@article{bergamino2024distinguishing,
   author = {Maurizio Bergamino and Molly M. McElvogue and Ashley M. Stokes},
   doi = {10.1002/ALZ.084580},
   issn = {1552-5279},
   issue = {S2},
   journal = {Alzheimer's \& Dementia},
   month = {12},
   pages = {e084580},
   publisher = {John Wiley \& Sons, Ltd},
   title = {Distinguishing Early Mild Cognitive Impairment from Late Mild Cognitive Impairment through Free-Water Diffusion Tensor Imaging: A Comparative Analysis},
   volume = {20},
   url = {https://onlinelibrary.wiley.com/doi/full/10.1002/alz.084580 https://onlinelibrary.wiley.com/doi/abs/10.1002/alz.084580 https://alz-journals.onlinelibrary.wiley.com/doi/10.1002/alz.084580},
   year = {2024}
}

@article{langhein2022association,
   author = {Mina Langhein and Johanna Seitz-Holland and Amanda E. Lyall and Ofer Pasternak and Natalia Chunga and Suheyla Cetin-Karayumak and Antoni Kubicki and Christoph Mulert and Randall T. Espinoza and Katherine L. Narr and Marek Kubicki},
   doi = {10.1016/J.JAD.2022.06.043},
   issn = {15732517},
   journal = {Journal of affective disorders},
   month = {10},
   pages = {78},
   pmid = {35779673},
   publisher = {Elsevier B.V.},
   title = {Association between peripheral inflammation and free-water imaging in Major Depressive Disorder before and after ketamine treatment – A pilot study},
   volume = {314},
   url = {https://pmc.ncbi.nlm.nih.gov/articles/PMC11186306/},
   year = {2022}
}

@article{li2025white,
   author = {Weicheng Li and Zerui You and Chengyu Wang and Xiaofeng Lan and Fan Zhang and Zhibo Hu and Xiaoyu Chen and Zhanjie Luo and Yexian Zeng and Yiying Chen and Yifang Chen and Siming Mai and Robin Shao and Hanna Lu and Roger S. McIntyre and Xiangdong Sun and Yuping Ning and Yanling Zhou},
   doi = {10.1038/S41398-025-03643-6},
   issn = {21583188},
   issue = {1},
   journal = {Translational Psychiatry},
   month = {12},
   pages = {407},
   pmid = {41107235},
   publisher = {Springer Nature},
   title = {White matter free water and depressive symptoms in medication-free depressed adolescents: moderation by peripheral inflammation},
   volume = {15},
   url = {https://pmc.ncbi.nlm.nih.gov/articles/PMC12534485/},
   year = {2025}
}

@article{roy2024quantifying,
   author = {Jean Charles Roy and Renaud Hédouin and Thomas Desmidt and Sébastien Dam and Iris Mirea-Grivel and Louise Weyl and Elise Bannier and Laurent Barantin and Dominique Drapier and Jean Marie Batail and Renaud David and Julie Coloigner and Gabriel H. Robert},
   doi = {10.1016/J.BPSC.2024.04.002},
   issn = {2451-9030},
   issue = {7},
   journal = {Biological psychiatry. Cognitive neuroscience and neuroimaging},
   month = {7},
   pages = {639-649},
   pmid = {38615911},
   publisher = {Biol Psychiatry Cogn Neurosci Neuroimaging},
   title = {Quantifying Apathy in Late-Life Depression: Unraveling Neurobehavioral Links Through Daily Activity Patterns and Brain Connectivity Analysis},
   volume = {9},
   url = {https://pubmed.ncbi.nlm.nih.gov/38615911/},
   year = {2024}
}

@article{kochunov2010fractional,
   author = {P. Kochunov and D. E. Williamson and J. Lancaster and P. Fox and J. Cornell and J. Blangero and D. C. Glahn},
   doi = {10.1016/J.NEUROBIOLAGING.2010.01.014},
   issn = {01974580},
   issue = {1},
   journal = {Neurobiology of aging},
   month = {1},
   pages = {9},
   pmid = {20122755},
   title = {Fractional anisotropy of water diffusion in cerebral white matter across the lifespan},
   volume = {33},
   url = {https://pmc.ncbi.nlm.nih.gov/articles/PMC2906767/},
   year = {2010}
}

@article{cousineau2017testretest,
   author = {Martin Cousineau and Pierre Marc Jodoin and Félix C. Morency and Verena Rozanski and Marilyn Grand'Maison and Barry J. Bedell and Maxime Descoteaux},
   doi = {10.1016/J.NICL.2017.07.020},
   issn = {22131582},
   journal = {NeuroImage : Clinical},
   pages = {222},
   pmid = {28794981},
   publisher = {Elsevier Inc.},
   title = {A test-retest study on Parkinson's PPMI dataset yields statistically significant white matter fascicles},
   volume = {16},
   url = {https://pmc.ncbi.nlm.nih.gov/articles/PMC5547250/},
   year = {2017}
}

@article{mullner2011modern,
   author = {Daniel Müllner},
   month = {9},
   title = {Modern hierarchical, agglomerative clustering algorithms},
   url = {https://arxiv.org/abs/1109.2378v1},
     journal={arXiv preprint arXiv:1109.2378},
   year = {2011}
   
}

@article{shahapure2020cluster,
   author = {Ketan Rajshekhar Shahapure and Charles Nicholas},
   doi = {10.1109/DSAA49011.2020.00096},
   journal = {Proceedings - 2020 IEEE 7th International Conference on Data Science and Advanced Analytics, DSAA 2020},
   month = {10},
   pages = {747-748},
   publisher = {Institute of Electrical and Electronics Engineers Inc.},
   title = {Cluster quality analysis using silhouette score},
   year = {2020}
}

@article{suarez2012automated,
   author = {Ralph O. Suarez and Olivier Commowick and Sanjay P. Prabhu and Simon K. Warfield},
   doi = {10.1016/J.NEUROIMAGE.2011.11.043},
   issn = {1053-8119},
   issue = {4},
   journal = {NeuroImage},
   month = {2},
   pages = {3690-3700},
   pmid = {22155046},
   publisher = {Academic Press},
   title = {Automated delineation of white matter fiber tracts with a multiple region-of-interest approach},
   volume = {59},
   year = {2012}
}

@article{arsigny2006logeuclidean,
   author = {Vincent Arsigny and Olivier Commowick and Xavier Pennec and Nicholas Ayache},
   doi = {10.1007/11866565_113/SAVE-RESEARCH},
   issn = {16113349},
   journal = {Lecture Notes in Computer Science (including subseries Lecture Notes in Artificial Intelligence and Lecture Notes in Bioinformatics)},
   pages = {924-931},
   pmid = {17354979},
   publisher = {Springer Verlag},
   title = {A log-euclidean framework for statistics on diffeomorphisms},
   volume = {4190 LNCS - I},
   url = {https://link.springer.com/chapter/10.1007/11866565_113},
   year = {2006}
}

@article{raz2006differential,
   author = {Naftali Raz and Karen M. Rodrigue},
   doi = {10.1016/J.NEUBIOREV.2006.07.001},
   issn = {01497634},
   issue = {6},
   journal = {Neuroscience and biobehavioral reviews},
   pages = {730},
   pmid = {16919333},
   title = {Differential aging of the brain: Patterns, cognitive correlates and modifiers},
   volume = {30},
   url = {https://pmc.ncbi.nlm.nih.gov/articles/PMC6601348/},
   year = {2006}
}

@article{xu2023metaanalysis,
   author = {Ellie P. Xu and Lynn Nguyen and Ellen Leibenluft and Jonathan P. Stange and Julia O. Linke},
   doi = {10.1017/S0033291723000107},
   issn = {1469-8978},
   issue = {7},
   journal = {Psychological medicine},
   month = {5},
   pages = {2721-2731},
   pmid = {37051913},
   publisher = {Psychol Med},
   title = {A meta-analysis on the uncinate fasciculus in depression},
   volume = {53},
   url = {https://pubmed.ncbi.nlm.nih.gov/37051913/},
   year = {2023}
}

@article{jiang2017microstructural,
   author = {Jing Jiang and You Jin Zhao and Xin Yu Hu and Ming Ying Du and Zi Qi Chen and Min Wu and Kai Ming Li and Hong Yan Zhu and Poornima Kumar and Qi Yong Gong},
   doi = {10.1503/JPN.150341},
   issn = {1488-2434},
   issue = {3},
   journal = {Journal of psychiatry \& neuroscience : JPN},
   month = {5},
   pages = {150-163},
   pmid = {27780031},
   publisher = {J Psychiatry Neurosci},
   title = {Microstructural brain abnormalities in medication-free patients with major depressive disorder: a systematic review and meta-analysis of diffusion tensor imaging},
   volume = {42},
   url = {https://pubmed.ncbi.nlm.nih.gov/27780031/},
   year = {2017}
}

@article{schmaal2020enigma,
   author = {Lianne Schmaal and Elena Pozzi and Tiffany C. Ho and Laura S. van Velzen and Ilya M. Veer and Nils Opel and Eus J.W. Van Someren and Laura K.M. Han and Lybomir Aftanas and André Aleman and Bernhard T. Baune and Klaus Berger and Tessa F. Blanken and Liliana Capitão and Baptiste Couvy-Duchesne and Kathryn R. Cullen and Udo Dannlowski and Christopher Davey and Tracy Erwin-Grabner and Jennifer Evans and Thomas Frodl and Cynthia H.Y. Fu and Beata Godlewska and Ian H. Gotlib and Roberto Goya-Maldonado and Hans J. Grabe and Nynke A. Groenewold and Dominik Grotegerd and Oliver Gruber and Boris A. Gutman and Geoffrey B. Hall and Ben J. Harrison and Sean N. Hatton and Marco Hermesdorf and Ian B. Hickie and Eva Hilland and Benson Irungu and Rune Jonassen and Sinead Kelly and Tilo Kircher and Bonnie Klimes-Dougan and Axel Krug and Nils Inge Landrø and Jim Lagopoulos and Jeanne Leerssen and Meng Li and David E.J. Linden and Frank P. MacMaster and Andrew M. McIntosh and David M.A. Mehler and Igor Nenadić and Brenda W.J.H. Penninx and Maria J. Portella and Liesbeth Reneman and Miguel E. Rentería and Matthew D. Sacchet and Philipp G. Sämann and Anouk Schrantee and Kang Sim and Jair C. Soares and Dan J. Stein and Leonardo Tozzi and Nic J.A. van Der Wee and Marie José van Tol and Robert Vermeiren and Yolanda Vives-Gilabert and Henrik Walter and Martin Walter and Heather C. Whalley and Katharina Wittfeld and Sarah Whittle and Margaret J. Wright and Tony T. Yang and Carlos Zarate and Sophia I. Thomopoulos and Neda Jahanshad and Paul M. Thompson and Dick J. Veltman},
   doi = {10.1038/s41398-020-0842-6},
   issn = {2158-3188},
   issue = {1},
   journal = {Translational Psychiatry 2020 10:1},
   month = {5},
   pages = {172-},
   pmid = {32472038},
   publisher = {Nature Publishing Group},
   title = {ENIGMA MDD: seven years of global neuroimaging studies of major depression through worldwide data sharing},
   volume = {10},
   url = {https://www.nature.com/articles/s41398-020-0842-6},
   year = {2020}
}

@article{disease2003unified,
  title={The unified Parkinson's disease rating scale (UPDRS): status and recommendations},
  author={Movement Disorder Society Task Force on Rating Scales for Parkinson's Disease},
  journal={Movement Disorders},
  volume={18},
  number={7},
  pages={738--750},
  year={2003},
  publisher={Wiley Online Library}
}

@article{mattis1976mental,
  title={Mental status examination for organic mental syndrome in the elderly patient},
  author={Mattis, Steve},
  journal={Geriatric psychiatry},
  year={1976},
  publisher={Grune and Stratton}
}

@article{lebihan2001diffusion,
  title={Diffusion tensor imaging: concepts and applications},
  author={Le Bihan, Denis and Mangin, Jean-Fran{\c{c}}ois and Poupon, Cyril and Clark, Chris A and Pappata, Sabina and Molko, Nicolas and Chabriat, Hughes},
  journal={Journal of Magnetic Resonance Imaging: An Official Journal of the International Society for Magnetic Resonance in Medicine},
  volume={13},
  number={4},
  pages={534--546},
  year={2001},
  publisher={Wiley Online Library}
}

@article{lindenmayer2003intersept,
   author = {J. P. Lindenmayer and Pal Czobor and Larry Alphs and Ann Marie Nathan and Ravi Anand and Zahur Islam and James C.Y. Chou and Saide Altinsan and Siemion Altman and Likiana Avigo and Richard Balon and Vanda Beněsová and Luis Bengochea and Alberto Bertoldi and Elisabeth Bokowska and Marc Bourgeois and Bernardo Carpiniello and James Chou and Guy Chouinard and Libor Chvila and Jean Dalery and Liliana Dell'Osso and Carl Eisdorfer and Robin A. Emsley and T. A. Fahy and Vera Folnegovic and Sophie Frangou and Pedro Gargoloff and Alberto Giannelli and Alan I. Green and Richard Greenberg and George T. Grossberg and George Hsu and Naveed Iqbal and Miro Jakovljevic and Richard C. Josiassen and Akos Kassaifarkas and Frederic Khidichian and Mary Ann Knesevich and Jack Krasuski and Veronica Larach and Michael Lesem and Pierre Michel Llorca and Jean Pierre Lindenmayer and Stephen Martin and Muriel Maurel-Raymondet and Herbert Meltzer and Laszlo Mod and Eva Morik and Carlos Morra and Ann Mortimer and Gyorgy Ostorharics-Horvath and Ivo Paclt and Jorg J. Pahl and Jeffrey Lee Peters and Rosario Piolo and Michael G. Plopper and Thomas Posever and Delbert Robinson and Carlo Andrea Robotti and Oladapo Tomori and Santha Vaidain and Zdèoka Vyhnándová and Marie Agathe Zimmerman},
   doi = {10.1016/S0920-9964(02)00335-3},
   issn = {09209964},
   issue = {1-2},
   journal = {Schizophrenia Research},
   month = {9},
   pages = {161-170},
   pmid = {12892870},
   publisher = {Elsevier},
   title = {The InterSePT scale for suicidal thinking reliability and validity},
   volume = {63},
   url = {https://pubmed.ncbi.nlm.nih.gov/12892870/},
   year = {2003}
}

@article{holtzmann2016quelle,
   author = {Jérôme Holtzmann and Raphaëlle Richieri and Ghassen Saba and Najib Allaïli and Rémy Bation and Fanny Moliere and Isabel Nieto and Frank Bellivier and Djamila Bennabi and Maxime Bubrovszky and Vincent Camus and Thomas Charpeaud and Pierre Courvoisier and Frédéric Haesebaert and Olivier Doumy and Philippe Courtet and Wissam El-Hage and Marion Garnier and Thierry d'Amato and Christophe Lançon and Marion Leboyer and Pierre Michel Llorca and Guillaume Vaiva and Thierry Bougerol and Bruno Aouizerate and Emmanuel Haffen},
   doi = {10.1016/J.LPM.2016.02.002},
   issn = {0755-4982},
   issue = {3},
   journal = {La Presse Medicale},
   month = {3},
   pages = {323-328},
   pmid = {26970938},
   publisher = {Elsevier Masson},
   title = {Quelle définition pour la dépression résistante ?},
   volume = {45},
   year = {2016}
}

@article{montgomery1979new,
  title={A new depression scale designed to be sensitive to change},
  author={Montgomery, Stuart A and {\AA}sberg, MARIE},
  journal={The British journal of psychiatry},
  volume={134},
  number={4},
  pages={382--389},
  year={1979},
  publisher={Cambridge University Press}
}

@inproceedings{corouge2004towards,
  title={Towards a shape model of white matter fiber bundles using diffusion tensor MRI},
  author={Corouge, Isabelle and Gouttard, Sylvain and Gerig, Guido},
  booktitle={2004 2nd IEEE international symposium on biomedical imaging: nano to macro (IEEE Cat No. 04EX821)},
  pages={344--347},
  year={2004},
  organization={IEEE}
}

@article{mishra2015toward,
   author = {Virendra Mishra and Xiaohu Guo and Mauricio R. Delgado and Hao Huang},
   doi = {10.1002/MRM.25548},
   issn = {1522-2594},
   issue = {6},
   journal = {Magnetic resonance in medicine},
   month = {12},
   pages = {1768-1779},
   pmid = {25447208},
   publisher = {Magn Reson Med},
   title = {Toward tract-specific fractional anisotropy (TSFA) at crossing-fiber regions with clinical diffusion MRI},
   volume = {74},
   url = {https://pubmed.ncbi.nlm.nih.gov/25447208/},
   year = {2015}
}

@article{panagiotaki2012compartment,
   author = {Eleftheria Panagiotaki and Torben Schneider and Bernard Siow and Matt G. Hall and Mark F. Lythgoe and Daniel C. Alexander},
   doi = {10.1016/J.NEUROIMAGE.2011.09.081},
   issn = {1095-9572},
   issue = {3},
   journal = {NeuroImage},
   month = {2},
   pages = {2241-2254},
   pmid = {22001791},
   publisher = {Neuroimage},
   title = {Compartment models of the diffusion MR signal in brain white matter: a taxonomy and comparison},
   volume = {59},
   url = {https://pubmed.ncbi.nlm.nih.gov/22001791/},
   year = {2012}
}

@article{zhang2018anatomically,
   author = {Fan Zhang and Ye Wu and Isaiah Norton and Laura Rigolo and Yogesh Rathi and Nikos Makris and Lauren J. O'Donnell},
   doi = {10.1016/J.NEUROIMAGE.2018.06.027},
   issn = {1053-8119},
   journal = {NeuroImage},
   month = {10},
   pages = {429-447},
   pmid = {29920375},
   publisher = {Academic Press},
   title = {An anatomically curated fiber clustering white matter atlas for consistent white matter tract parcellation across the lifespan},
   volume = {179},
   year = {2018}
}

@article{scherrer2012parametric,
   author = {Benoit Scherrer and Simon K. Warfield},
   doi = {10.1371/JOURNAL.PONE.0048232},
   issn = {1932-6203},
   issue = {11},
   journal = {PLOS ONE},
   month = {11},
   pages = {e48232},
   pmid = {23189128},
   publisher = {Public Library of Science},
   title = {Parametric Representation of Multiple White Matter Fascicles from Cube and Sphere Diffusion MRI},
   volume = {7},
   url = {https://journals.plos.org/plosone/article?id=10.1371/journal.pone.0048232},
   year = {2012}
}

@article{sheehan1994mini,
  title={MINI: Mini International Neuropsychiatric Interview},
  author={Sheehan, D and Janavs, J and Baker, R and Harnett-Sheehan, K and Knapp, E and Sheehan, M and others},
  journal={English version},
  volume={6},
  number={0},
  year={1994}
}

@inproceedings{tournier2010improved,
  title={Improved probabilistic streamlines tractography by 2nd order integration over fibre orientation distributions},
  author={Tournier, J Donald and Calamante, Fernando and Connelly, Alan and others},
  booktitle={Proceedings of the international society for magnetic resonance in medicine},
  number={10.1016},
  year={2010},
  organization={Stockholm}
}

@article{tournier2019mrtrix,
   author = {J. Donald Tournier and Robert Smith and David Raffelt and Rami Tabbara and Thijs Dhollander and Maximilian Pietsch and Daan Christiaens and Ben Jeurissen and Chun Hung Yeh and Alan Connelly},
   doi = {10.1016/J.NEUROIMAGE.2019.116137},
   issn = {1053-8119},
   journal = {NeuroImage},
   month = {11},
   pages = {116137},
   pmid = {31473352},
   publisher = {Academic Press},
   title = {MRtrix3: A fast, flexible and open software framework for medical image processing and visualisation},
   volume = {202},
   year = {2019}
}

@article{jeurissen2014multitissue,
   author = {Ben Jeurissen and Jacques Donald Tournier and Thijs Dhollander and Alan Connelly and Jan Sijbers},
   doi = {10.1016/J.NEUROIMAGE.2014.07.061},
   issn = {1053-8119},
   journal = {NeuroImage},
   month = {12},
   pages = {411-426},
   pmid = {25109526},
   publisher = {Academic Press},
   title = {Multi-tissue constrained spherical deconvolution for improved analysis of multi-shell diffusion MRI data},
   volume = {103},
   year = {2014}
}

@article{raffelt2017investigating,
   author = {David A. Raffelt and J. Donald Tournier and Robert E. Smith and David N. Vaughan and Graeme Jackson and Gerard R. Ridgway and Alan Connelly},
   doi = {10.1016/J.NEUROIMAGE.2016.09.029},
   issn = {1053-8119},
   journal = {NeuroImage},
   month = {1},
   pages = {58-73},
   pmid = {27639350},
   publisher = {Academic Press},
   title = {Investigating white matter fibre density and morphology using fixel-based analysis},
   volume = {144},
   year = {2017}
}

@article{zheng2026agfstractometry,
   author = {Ruixi Zheng and Wei Zhang and Yijie Li and Xi Zhu and Zhou Lan and Jarrett Rushmore and Yogesh Rathi and Nikos Makris and Lauren J. O’Donnell and Fan Zhang},
   doi = {10.1016/J.MEDIA.2025.103892},
   issn = {1361-8415},
   journal = {Medical Image Analysis},
   month = {3},
   pages = {103892},
   pmid = {41380339},
   publisher = {Elsevier},
   title = {AGFS-tractometry: A novel atlas-guided fine-scale tractometry approach for enhanced along-tract group statistical comparison using diffusion MRI tractography},
   volume = {109},
   url = {https://www.sciencedirect.com/science/article/pii/S1361841525004384},
   year = {2026}
}

@article{yeatman2012tract,
   author = {Jason D. Yeatman and Robert F. Dougherty and Nathaniel J. Myall and Brian A. Wandell and Heidi M. Feldman},
   doi = {10.1371/JOURNAL.PONE.0049790},
   issn = {1932-6203},
   issue = {11},
   journal = {PloS one},
   month = {11},
   pmid = {23166771},
   publisher = {PLoS One},
   title = {Tract profiles of white matter properties: automating fiber-tract quantification},
   volume = {7},
   url = {https://pubmed.ncbi.nlm.nih.gov/23166771/},
   year = {2012}
}

@article{chandio2023buan,
   author = {Bramsh Qamar Chandio and Sophia Thomopoulos and Paul Thompson and Jaroslaw Harezlak and Eleftherios Garyfallidis},
   doi = {10.58530/2023/3965},
   journal = {2023 ISMRM \& ISMRT Annual Meeting},
   month = {5},
   publisher = {ISMRM},
   title = {BUAN 2.0, streamlines as functions, nonlinear registration, and subdivision of bundles for advanced tractometry},
   year = {2023}
}

@article{neher2023radiomic,
   author = {Peter Neher and Dusan Hirjak and Klaus Maier-Hein},
   doi = {10.21203/RS.3.RS-2950610/V1},
   institution = {Research Square},
   journal = {Research Square},
   month = {5},
   pages = {rs.3.rs-2950610},
   pmid = {37292645},
   title = {Radiomic tractometry: a rich and tract-specific class of imaging biomarkers for neuroscience and medical applications},
   url = {https://pmc.ncbi.nlm.nih.gov/articles/PMC10246281/},
   year = {2023}
}

@article{garyfallidis2012quickbundles,
   author = {Eleftherios Garyfallidis and Matthew Brett and Marta Morgado Correia and Guy B. Williams and Ian Nimmo-Smith},
   doi = {10.3389/FNINS.2012.00175},
   journal = {Frontiers in Neuroscience},
   pages = {175},
   pmid = {23248578},
   publisher = {Frontiers Media SA},
   title = {QuickBundles, a Method for Tractography Simplification},
   volume = {6},
   url = {https://pmc.ncbi.nlm.nih.gov/articles/PMC3518823/},
   year = {2012}
}

@article{eiter1994computing,
  title={Computing discrete Frechet distance},
  author={Eiter, Thomas and Mannila, Heikki},
  year={1994},  
  journal={Technical Report CD-TR 94 64, Christian Doppler Laboratory for Expert}
}

@article{hedouin2024microstructural,
	year = {2024},
	title = {Microstructural brain assessment in late-life depression and apathy using diffusion {MRI} multi-compartments models and tractometry},
	volume = {14},
	issn = {2045-2322},
	url = {https://www.nature.com/articles/s41598-024-67535-3},
	doi = {10.1038/s41598-024-67535-3},
	pages = {18193},
	number = {1},
	journal = {Scientific Reports},
	publisher = {Nature Publishing Group},
	author = {Hédouin, Renaud and Roy, Jean-Charles and Desmidt, Thomas and Robert, Gabriel and Coloigner, Julie},
	date = {2024-08-06},
}

@article{chandio2020bundle,
	year = {2020},
	title = {Bundle analytics, a computational framework for investigating the shapes and profiles of brain pathways across populations},
	volume = {10},
	issn = {2045-2322},
	url = {https://www.nature.com/articles/s41598-020-74054-4},
	doi = {10.1038/s41598-020-74054-4},
	pages = {17149},
	number = {1},
	journal = {Scientific Reports},
	publisher = {Nature Publishing Group},
	author = {Chandio, Bramsh Qamar and Risacher, Shannon Leigh and Pestilli, Franco and Bullock, Daniel and Yeh, Fang-Cheng and Koudoro, Serge and Rokem, Ariel and Harezlak, Jaroslaw and Garyfallidis, Eleftherios},
	date = {2020-10-13},
}

@article{wasserthal2018tractseg,
	title = {{TractSeg} - Fast and accurate white matter tract segmentation},
	volume = {183},
	issn = {1053-8119},
   year= {2018},
	url = {https://www.sciencedirect.com/science/article/pii/S1053811918306864},
	doi = {10.1016/j.neuroimage.2018.07.070},
	pages = {239--253},
	journal = {{NeuroImage}},
	author = {Wasserthal, Jakob and Neher, Peter and Maier-Hein, Klaus H.},
	date = {2018-12-01},
}

@misc{st-onge2023bundleseg,
	year = {2023},
	title = {{BundleSeg}: A versatile, reliable and reproducible approach to white matter bundle segmentation},
	url = {http://arxiv.org/abs/2308.10958},
	doi = {10.48550/arXiv.2308.10958},
	number = {{arXiv}:2308.10958},
	publisher = {{arXiv}},
	author = {St-Onge, Etienne and Schilling, Kurt G. and Rheault, Francois},
	date = {2023-08-21},
}

@article{guimond2000average,
	year = {2000},
	title = {Average Brain Models: A Convergence Study},
	volume = {77},
	issn = {1077-3142},
	url = {https://www.sciencedirect.com/science/article/pii/S1077314299908159},
	doi = {10.1006/cviu.1999.0815},
	pages = {192--210},
	number = {2},
	journal = {Computer Vision and Image Understanding},
	author = {Guimond, Alexandre and Meunier, Jean and Thirion, Jean-Philippe},
	date = {2000-02-01},
}

@misc{jakob2018high,
	title = {High quality white matter reference tracts},
	url = {https://zenodo.org/records/1477956},
	doi = {10.5281/zenodo.1477956},
	version = {1.2.0},
	publisher = {Zenodo},
	author = {Jakob, Wasserthal and Peter, Neher and Klaus, Maier-Hein},
	year = {2018},
	date = {2018-11-05},
}

@article{durantel2025riemannian,
	title = {A Riemannian framework for incorporating white matter bundle prior in orientation distribution function based tractography algorithms},
	volume = {20},
	issn = {1932-6203},
	url = {https://journals.plos.org/plosone/article?id=10.1371/journal.pone.0304449},
	doi = {10.1371/journal.pone.0304449},
	pages = {e0304449},
	number = {3},
   year = {2025},
	journal = {{PLOS} {ONE}},
	publisher = {Public Library of Science},
	author = {Durantel, Thomas and Girard, Gabriel and Caruyer, Emmanuel and Commowick, Olivier and Coloigner, Julie},
	date = {2025-03-25},
}

@article{chamberland2019dimensionality,
	year = {2019},
	title = {Dimensionality reduction of diffusion {MRI} measures for improved tractometry of the human brain},
	volume = {200},
	issn = {1053-8119},
	url = {https://www.sciencedirect.com/science/article/pii/S1053811919305051},
	doi = {10.1016/j.neuroimage.2019.06.020},
	pages = {89--100},
	journal = {{NeuroImage}},
	author = {Chamberland, Maxime and Raven, Erika P. and Genc, Sila and Duffy, Kate and Descoteaux, Maxime and Parker, Greg D. and Tax, Chantal M. W. and Jones, Derek K.},
	date = {2019-10-15},
}

@article{koshiyama2019white,
	year = {2019},
  title = {White Matter Microstructural Alterations across Four Major Psychiatric Disorders: Mega-Analysis Study in 2937 Individuals},
  author = {Koshiyama, Daisuke and Fukunaga, Masaki and Okada, Naohiro and Morita, Kentaro and Nemoto, Kiyotaka and Usui, Kaori and Yamamori, Hidenaga and Yasuda, Yuka and Fujimoto, Michiko and Kudo, Noriko and Azechi, Hirotsugu and Watanabe, Yoshiyuki and Hashimoto, Naoki and Narita, Hisashi and Kusumi, Ichiro and Ohi, Kazutaka and Shimada, Takamitsu and Kataoka, Yuzuru and Yamamoto, Maeri and Ozaki, Norio and Okada, Go and Okamoto, Yasumasa and Harada, Kenichiro and Matsuo, Koji and Yamasue, Hidenori and Abe, Osamu and Hashimoto, Ryuichiro and Takahashi, Tsutomu and Hori, Tomoki and Nakataki, Masahito and Onitsuka, Toshiaki and Holleran, Laurena and Jahanshad, Neda and family=Erp, given=Theo G.M., prefix=van, useprefix=true and Turner, Jessica and Donohoe, Gary and Thompson, Paul M. and Kasai, Kiyoto and Hashimoto, Ryota},
  date = {2019-11},
  journal = {Molecular Psychiatry 2019 25:4},
  volume = {25},
  number = {4},
  pages = {883--895},
  publisher = {Nature Publishing Group},
  issn = {14765578},
  doi = {10.1038/s41380-019-0553-7},
  url = {https://www.nature.com/articles/s41380-019-0553-7}
}

@article{kruper2021evaluating,
	year = {2021},
  title = {Evaluating the Reliability of Human Brain White Matter Tractometry},
  author = {Kruper, John and Yeatman, Jason D. and Richie-Halford, Adam and Bloom, David and Grotheer, Mareike and Caffarra, Sendy and Kiar, Gregory and Karipidis, Iliana I. and Roy, Ethan and Chandio, Bramsh Q. and Garyfallidis, Eleftherios and Rokem, Ariel},
  date = {2021-11},
  journal = {Aperture neuro},
  volume = {1},
  number = {1},
  publisher = {Apert Neuro},
  issn = {2957-3963},
  doi = {10.52294/e6198273-b8e3-4b63-babb-6e6b0da10669},
  url = {https://pubmed.ncbi.nlm.nih.gov/35079748/}
}

@article{liao2012dysfunction,
	year = {2012},
  title = {Dysfunction of Fronto-Limbic Brain Circuitry in Depression},
  author = {Liao, C. and Feng, Z. and Zhou, D. and Dai, Q. and Xie, B. and Ji, B. and Wang, X. and Wang, X.},
  date = {2012-01},
  journal = {Neuroscience},
  volume = {201},
  pages = {231--238},
  publisher = {Neuroscience},
  issn = {03064522},
  doi = {10.1016/j.neuroscience.2011.10.053},
  url = {https://pubmed.ncbi.nlm.nih.gov/22119640/}
}

@article{mesbah2023association,
	year = {2023},
  title = {Association between the Fronto-Limbic Network and Cognitive and Emotional Functioning in Individuals with Bipolar Disorder: A Systematic Review and Meta-Analysis},
  author = {Mesbah, Rahele and Koenders, Manja A. and Wee, Nic J.A. Van Der and Giltay, Erik J. and Hemert, Albert M. Van and Leeuw, Max De},
  date = {2023-05},
  journal = {JAMA Psychiatry},
  volume = {80},
  number = {5},
  pages = {432--440},
  publisher = {American Medical Association},
  issn = {2168622X},
  doi = {10.1001/jamapsychiatry.2023.0131},
  url = {https://jamanetwork.com/journals/jamapsychiatry/fullarticle/2802944}
}

@article{sumra2025regional,
	year = {2025},
  title = {Regional Free-Water Diffusion Is More Strongly Related to Neuroinflammation than Neurodegeneration},
  author = {Sumra, Vishaal and Hadian, Mohsen and Dilliott, Allison A. and Farhan, Sali M.K. and Frank, Andrew R. and Lang, Anthony E. and Roberts, Angela C. and Troyer, Angela and Arnott, Stephen R. and Marras, Connie and Tang-Wai, David F. and Finger, Elizabeth and Rogaeva, Ekaterina and Orange, Joseph B. and Ramirez, Joel and Zinman, Lorne and Binns, Malcolm and Borrie, Michael and Freedman, Morris and Ozzoude, Miracle and Bartha, Robert and Swartz, Richard H. and Munoz, David and Masellis, Mario and Black, Sandra E. and Dixon, Roger A. and Dowlatshahi, Dar and Grimes, David and Hassan, Ayman and Hegele, Robert A. and Kumar, Sanjeev and Pasternak, Stephen and Pollock, Bruce and Rajji, Tarek and Sahlas, Demetrios and Saposnik, Gustavo and Tartaglia, Maria Carmela},
  date = {2025-06},
  journal = {Journal of Neurology 2025 272:7},
  volume = {272},
  number = {7},
  pages = {478-},
  publisher = {Springer},
  issn = {14321459},
  doi = {10.1007/s00415-025-13201-1},
  url = {https://link.springer.com/article/10.1007/s00415-025-13201-1}
}

@article{wijtenburg2012relationship,
	year = {2012},
  title = {Relationship between Fractional Anisotropy of Cerebral White Matter and Metabolite Concentrations Measured Using {{1H}} Magnetic Resonance Spectroscopy in Healthy Adults},
  author = {Wijtenburg, S. A. and McGuire, S. A. and Rowland, L. M. and Sherman, P. M. and Lancaster, J. L. and Tate, D. F. and Hardies, L. J. and Patel, B. and Glahn, D. C. and Hong, L. E. and Fox, P. T. and Kochunov, P.},
  date = {2012-02},
  journal = {NeuroImage},
  volume = {0},
  pages = {161},
  issn = {10538119},
  doi = {10.1016/j.neuroimage.2012.10.014},
  url = {https://pmc.ncbi.nlm.nih.gov/articles/PMC3779655/}
}

\pagebreak
\setcounter{section}{0}
\setcounter{figure}{0}
\setcounter{table}{0}
\setcounter{equation}{0}

\renewcommand{\thesection}{S\arabic{section}}
\renewcommand{\thefigure}{S\arabic{figure}}
\renewcommand{\thetable}{S\arabic{table}}
\renewcommand{\theequation}{S\arabic{equation}}
% \begin{document}

\section{Demographic statistics}

\begin{table}[h]
    \centering
    \caption{Demographic characteristics of the LLD cohort. The table reports the number of subjects and mean age (with standard deviation) for each clinical group: healthy controls and subjects with LLD. The significance levels of the \textit{Welch}'s t-test is showwn in the $p$-value row.}
    \label{tab:actidep_demographics}
    \begin{tabular}{lcc}
\toprule
Group & N. subjects & Age \\
\midrule
Control & 24 & 74.6 $\pm$ 5.2 \\
LLD & 36 & 75.2 $\pm$ 6.5 \\
\bottomrule
\end{tabular}

\end{table}

\begin{table}[h]
    \centering
    \caption{Comparability in age between healthy controls, remitted and treatment-resistant subjects of the ELD cohort. The significance levels of the \textit{Welch}'s ANOVA is showwn in the $p$-value row.}
    \label{tab:amynet_demographics}
    \begin{tabular}{lcc}
\toprule
Group & N. subjects & Age \\
\midrule
Remission & 25 & 24.7 $\pm$ 5.4 \\
Resistant & 25 & 25.6 $\pm$ 5.0 \\
Control & 24 & 28.1 $\pm$ 6.0 \\
\bottomrule
\end{tabular}

\end{table}

\begin{table}[h]
    \centering
    \caption{Comparability in age, MADRS and duration of depression between remitted and treatment-resistant subjects of the ELD cohort. The significance levels of the t-tests are shown in the $p$-value row.}
    \begin{tabular}{lcccc}
\toprule
Group & N & Age (years) & MADRS & Depression duration (months) \\
\midrule
Remission & 25 & 24.7 $\pm$ 5.4 & 5.6 $\pm$ 3.9 & 51.0 $\pm$ 41.9 \\
Resistant & 25 & 25.6 $\pm$ 5.0 & 26.2 $\pm$ 5.9 & 77.4 $\pm$ 74.0 \\
\midrule
\textit{p}-value &  & 0.551 & \textbf{0.000} & 0.132 \\
\bottomrule
\end{tabular}

    \label{tab:placeholder}
\end{table}
\pagebreak

\section{Point-level association}

\begin{figure}[!htbp]
    \centering
    \includegraphics[width=0.9\textwidth]{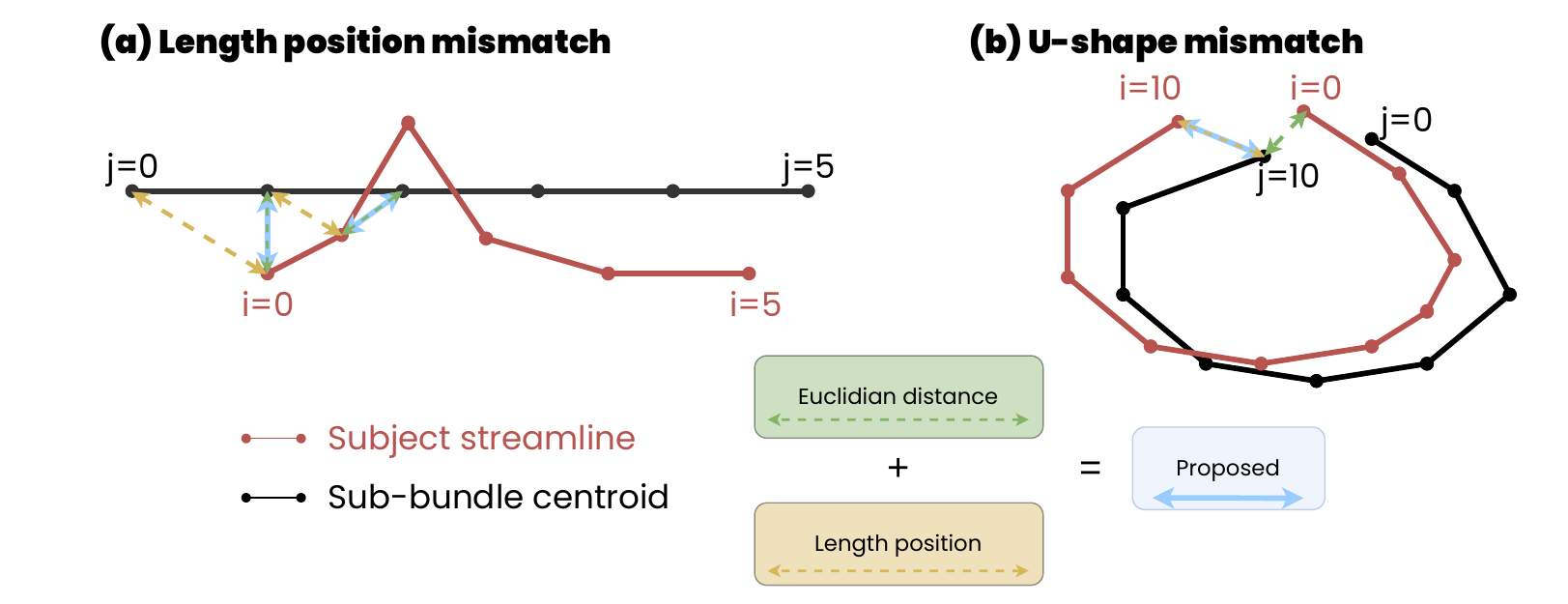}
    \caption{Illustration of the point-level association process between subject streamlines and cluster points. Each streamline point is embedded into an augmented space that jointly encodes spatial coordinates and normalized index position, allowing for a balanced association that penalizes both spatial displacement and length mismatch. The nearest cluster point is identified for each streamline point based on this composite distance, enabling the aggregation of microstructural metrics at anatomically corresponding locations along the bundle.}
    \label{fig:association_method}
\end{figure}

\pagebreak
\section{Incomplete sub-bundle associations}

\begin{figure}[h]
  \centering
  \begin{subfigure}{0.31\textwidth}
    \centering
    \resizebox{\linewidth}{!}{%
    \begin{tabular}{@{}lcccc@{}}
    \toprule
    Bundle                        & Centroid & Miss.~dep. & Miss.~HC &  Tot.~miss. \\ \midrule
    \multirow{2}{*}{CA}           & 0        & 3                & 1             & 4             \\
                                  & 1        & 3                & 1             & 4             \\ \midrule
    \multirow{2}{*}{FX left}      & 0        & 1                & 0             & 1             \\
                                  & 1        & 5                & 1             & 6             \\ \midrule
    \multirow{2}{*}{FX right}     & 0        & 3                & 2             & 5             \\
                                  & 1        & 1                & 1             & 2             \\ \midrule
    \multirow{3}{*}{ILF left}     & 0        & 4                & 5             & 9             \\
                                  & 1        & 2                & 2             & 4             \\
                                  & 2        & 2                & 3             & 5             \\ \midrule
    \multirow{3}{*}{ILF right}    & 0        & 6                & 1             & 7             \\
                                  & 1        & 1                & 0             & 1             \\
                                  & 2        & 2                & 0             & 2             \\ \midrule
    \multirow{2}{*}{SLF III left} & 0        & 1                & 1             & 2             \\
                                  & 1        & 0                & 1             & 1             \\ \midrule
    \multirow{2}{*}{SLF II right} & 0        & 2                & 5             & 7             \\
                                  & 1        & 1                & 0             & 1             \\ \midrule
    \multirow{2}{*}{SLF I left}   & 0        & 1                & 0             & 1             \\
                                  & 1        & 1                & 0             & 1             \\ \bottomrule 
    \end{tabular}%
    }
    \caption{}
  \end{subfigure}
  \hfill
  \begin{subfigure}{0.66\textwidth}
    \centering
    \includegraphics[width=\linewidth]{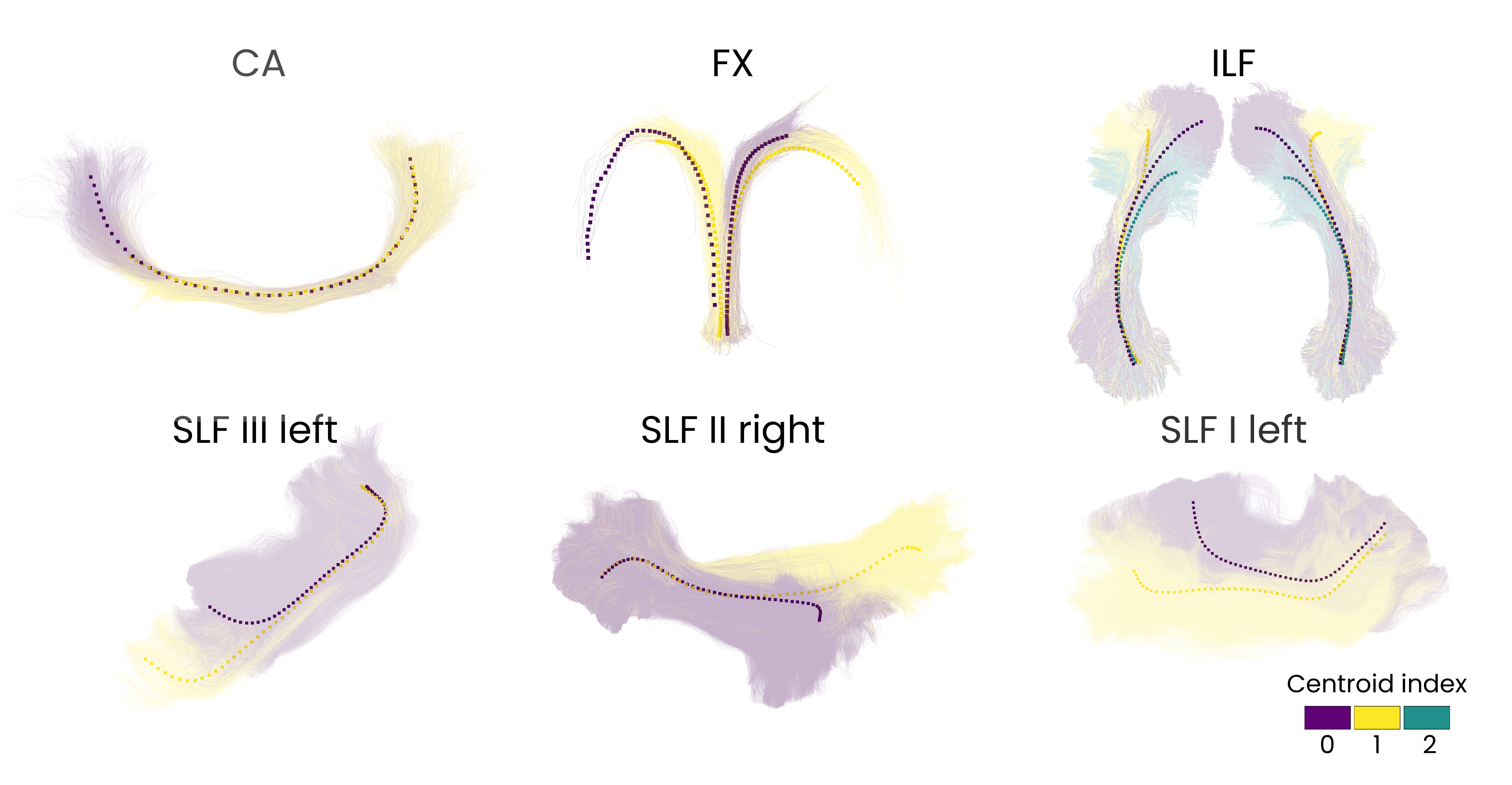}
    \caption{}
  \end{subfigure}
  
  \caption{(a) Missing association by centroid and per LLD group for excluded bundles (no complete sub-bundle), alongside an illustration (b) of the excluded bundles with color-coded sub-bundles and their respective centroids.}
  \label{fig:excluded_bundles}
\end{figure}

\pagebreak
\section{Bundle profiles}

\begin{figure}[h]
    \centering
    \includegraphics[width=0.85\textwidth]{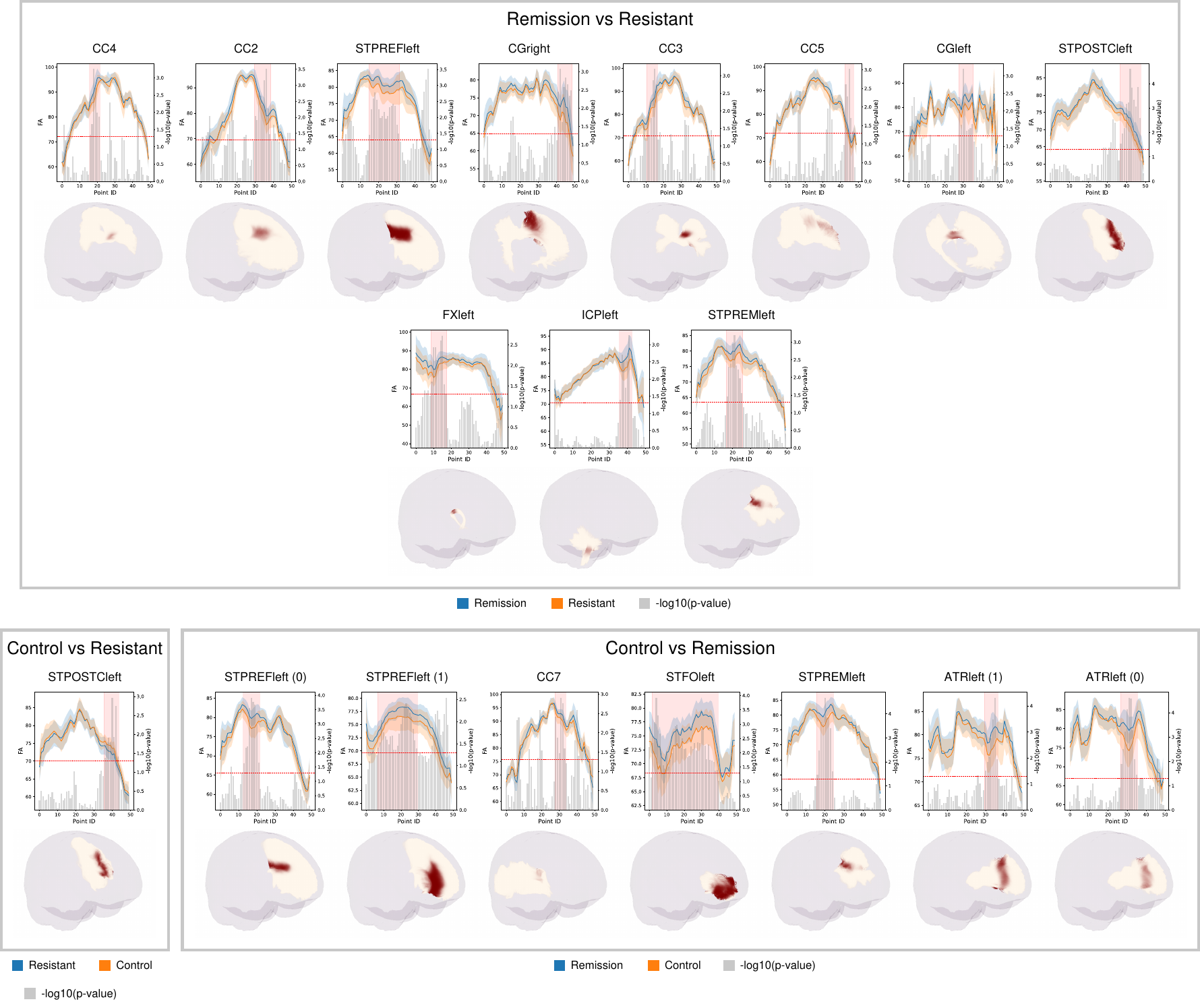}
    \caption{Along-tract FA profiles for significant sub-bundles in ELD group comparison analysis. The (arbitrary) number of the sub-bundle is indicated in parentheses when the same bundle appears multiple times. Each panel shows the group mean profiles with SD as shaded areas, and the point-wise $-10\log(p\text{-values})$ from the group comparison tests. Significant point clusters (multiple comparison) are highlighted in red.}
    \label{fig:along_tract_amynet_FA}
\end{figure}

\begin{figure}[h]
    \centering
    \includegraphics[width=0.99\textwidth]{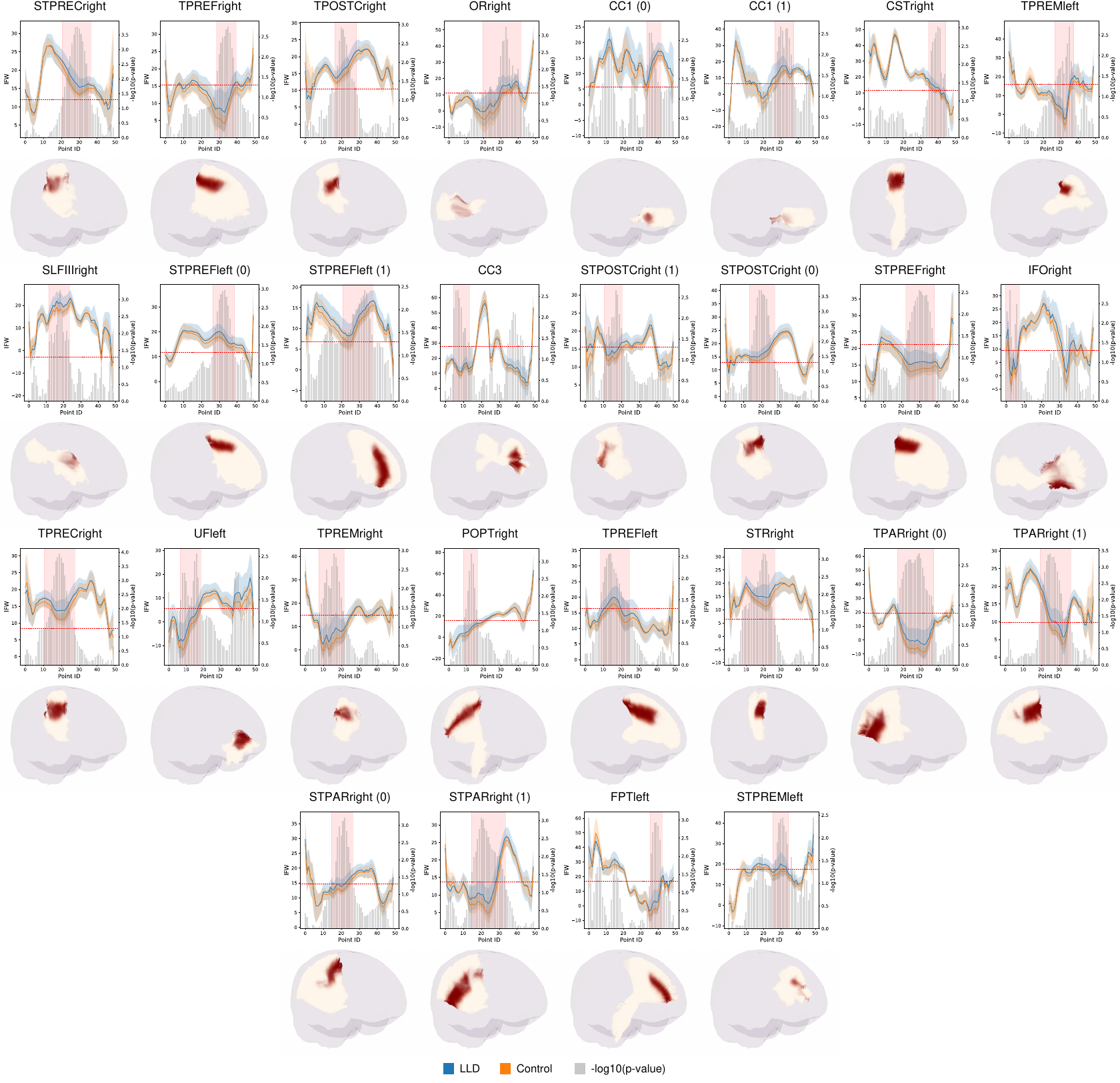}
    \caption{Along-tract IFW profiles for significant sub-bundles in LLD group comparison analysis. The (arbitrary) number of the sub-bundle is indicated in parentheses when the same bundle appears multiple times. Each panel shows the group mean profiles with SD as shaded areas, and the point-wise $-10\log(p\text{-values})$ from the group comparison tests. Significant point clusters (multiple comparison) are highlighted in red.}
    \label{fig:along_tract_actidep_IFW}
\end{figure}

\begin{figure}[h]
    \centering
    \includegraphics[width=0.75\textwidth]{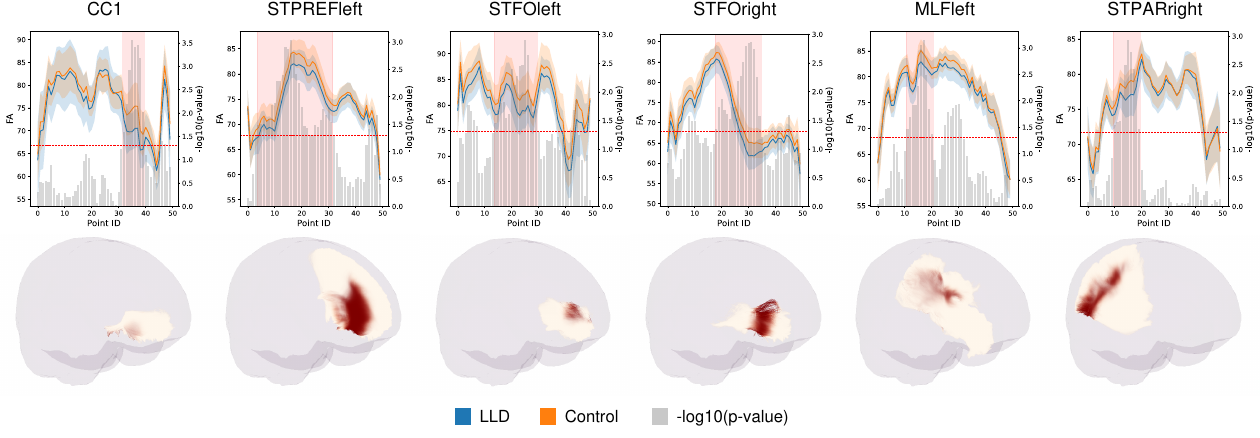}
    \caption{Along-tract FA profiles for significant sub-bundles in ELD group comparison analysis. The (arbitrary) number of the sub-bundle is indicated in parentheses when the same bundle appears multiple times. Each panel shows the group mean profiles with SD as shaded areas, and the point-wise $-10\log(p\text{-values})$ from the group comparison tests. Significant point clusters (multiple comparison) are highlighted in red.}
    \label{fig:along_tract_actidep_FA}
\end{figure}

% \end{document}
\end{document}